\documentclass[eprint,superscriptaddress,reprint]{revtex4-2}
\usepackage{float}
\usepackage{graphicx}
\usepackage{xcolor}
\usepackage{amsmath}
\usepackage{multirow}

\usepackage{times} 
\usepackage{color} 
\usepackage{comment} 
\usepackage{braket} 
\usepackage{esint} 
\usepackage{amsfonts}
\usepackage{amssymb}
\usepackage{bm}
\usepackage{dcolumn}

\usepackage[colorlinks,linkcolor=blue,anchorcolor=blue,urlcolor=blue,urlcolor=blue,citecolor=blue]{hyperref}

\usepackage{soul}

\begin{document}

\title{Fast entangling gates on fluxoniums via parametric modulation of plasmon interaction}

\author{Peng Zhao}
\email{shangniguo@sina.com}
\affiliation{Quantum Science Center of Guangdong-Hong Kong-Macao Greater Bay Area, Shenzhen 518045, China}
\author{Peng Xu}
\affiliation{Institute of Quantum Information and Technology,
Nanjing University of Posts and Telecommunications, Nanjing, Jiangsu 210003, China}
\affiliation{State Key Laboratory of Quantum Optics Technologies and Devices, Shanxi University, Taiyuan, 030006, China}
\author{Zheng-Yuan Xue}
\email{zyxue83@163.com}
\affiliation{Key Laboratory of Atomic and Subatomic Structure and Quantum Control (Ministry of Education), Guangdong Basic Research Center of Excellence for Structure and Fundamental Interactions of Matter, and School of Physics, South China Normal University, Guangzhou 510006, China}
\affiliation{Guangdong Provincial Key Laboratory of Quantum Engineering and Quantum Materials, Guangdong-Hong Kong Joint Laboratory of Quantum Matter, and Frontier Research Institute for Physics, South China Normal University, Guangzhou 510006, China}
\affiliation{Quantum Science Center of Guangdong-Hong Kong-Macao Greater Bay Area, Shenzhen 518045, China}
\date{\today}

\begin{abstract}

In superconducting quantum processors, exploring diverse control methods could offer essential 
versatility and redundancy to mitigate challenges such as frequency crowding, spurious couplings, control crosstalk, and fabrication variability, thus leading to better system-level performance. Here we introduce a control strategy for fast entangling gates in a scalable fluxonium architecture, utilizing parametric modulation of the plasmon interaction. In this architecture, fluxoniums are coupled via a tunable coupler, whose transition frequency is flux-modulated to control the inter-fluxonium plasmon interaction. A bSWAP-type interaction is activated by parametrically driving the coupler at the sum frequency of the plasmon transitions of the two fluxoniums, resulting in the simultaneous excitation or de-excitation of both plasmon modes. This strategy therefore allow the transitions between computational states and non-computational plasmon states, enabling the accumulation of conditional phases on the computational subspace and facilitating the realization of controlled-phase gates. By focusing on a specific case of these bSWAP-type interactions, we show that a simple drive pulse enables sub-100ns CZ gates with an intrinsic error below $10^{-4}$. Given its operational flexibility and extensibility, this approach could potentially offer a foundational framework for developing scalable fluxonium-based quantum processors.

\end{abstract}

\maketitle


\section{Introduction}\label{SecI}

Unlike the well-studied transmon qubit~\cite{Koch2007}, the fluxonium qubit~\cite{Manucharyan2009} exhibits a
complex, strongly anharmonic energy structure. Besides the qubit transition, which typically occurs at
frequencies around 100 MHz, the fluxonium also exhibits multiple accessible plasmon transitions spanning a broad range from a few gigahertz to over 10 GHz. On one hand, these plasmon modes in fluxonium systems increases susceptibility
to unintended activations of these transitions and spurious inter-mode couplings among fluxonium qubits and
ancillary circuits such as readout resonators and couplers, during quantum operations. On the other hand, this
spectral richness could provide the essential opportunity to mitigate challenges including frequency
crowding~\cite{Brink2018,Chen2014,Kelly2015}, spurious couplings~\cite{Chen2014,Kelly2015,Mundada2019,Muller2019}, control crosstalk~\cite{Barends2014}, and fabrication variability~\cite{Kreikebaum2020,Hertzberg2021,Pappas2024}, which is particularly critical in large-scale superconducting quantum processors. In this context, these transitions can, and indeed should, be carefully allocated to support distinct quantum operations such as two-qubit gates~\cite{Nesterov2018,Ficheux2021,Xiong2022,Simakov2023,Ding2023,Rosenfeld2024,Xiong2025,Zhao2025,Singh2025}, readout~\cite{Zhu2013,Lin2018,Nguyen2022,Stefanski2024,Bothara2025,Watanabe2025}, and
initialization~\cite{Watanabe2025,Manucharyan2009b,Zhang2021,Wang2024a}, thereby advancing the development
of scalable high-performance fluxonium-based quantum processors. Generally, without
detailed consideration of specific qubit characteristics, a potential allocation strategy may seek to maximize
frequency separation between bands dedicated to distinct operations.

For fluxonium qubits (the lowest five states labeled by $\{|0\rangle, |1\rangle, |2\rangle, |3\rangle, |4\rangle\}$), which are typically biased at the half-flux-quantum sweet spot to achieve high coherence times~\cite{Nguyen2019,Somoroff2023,Wang2025}, the small electric dipole moment of the qubit transition $|0\rangle\rightarrow |1\rangle$ complicates its use in quantum operations beyond single-qubit gates~\cite{Moskalenko2021,Moskalenko2022}, particularly within large-scale systems. However, among accessible higher-energy transitions, three plasmon transitions, namely $|0\rangle \rightarrow |3\rangle$, $|1\rangle \rightarrow |2\rangle$, and $|1\rangle \rightarrow |4\rangle$, possess transmon-like dipole moments, making them particularly relevant for implementing various quantum operations. Specifically, these transitions can facilitate strong dispersive coupling with readout resonators, enabling fast, high-fidelity qubit readout~\cite{Stefanski2024,Bothara2025}. They can also, in principle, be leveraged to realize microwave-activated controlled-phase gates by facilitating transitions between computational states and non-computational states (involving the excitation of these plasmon modes~\cite{Nesterov2018} or coupler modes~\cite{Simakov2023}). Apparently, both types of operations could inevitably cause leakage out of the computational subspace, presenting a significant challenge for quantum error correction~\cite{Miao2023}. Addressing this issue requires the incorporation of leakage removal operations, typically involving engineered interactions between fluxonium plasmons and a dissipative environment such as the readout resonator~\cite{Wang2024a}. It is therefore evident that the practical utilization of plasmon modes in
large-scale fluxonium-based quantum processors necessitates a flexible and scalable control strategy for
engineering plasmon interactions~\cite{Zhao2025,Zhao2025b}.

\begin{figure}[tbp]
\begin{center}
\includegraphics[keepaspectratio=true,width=\columnwidth]{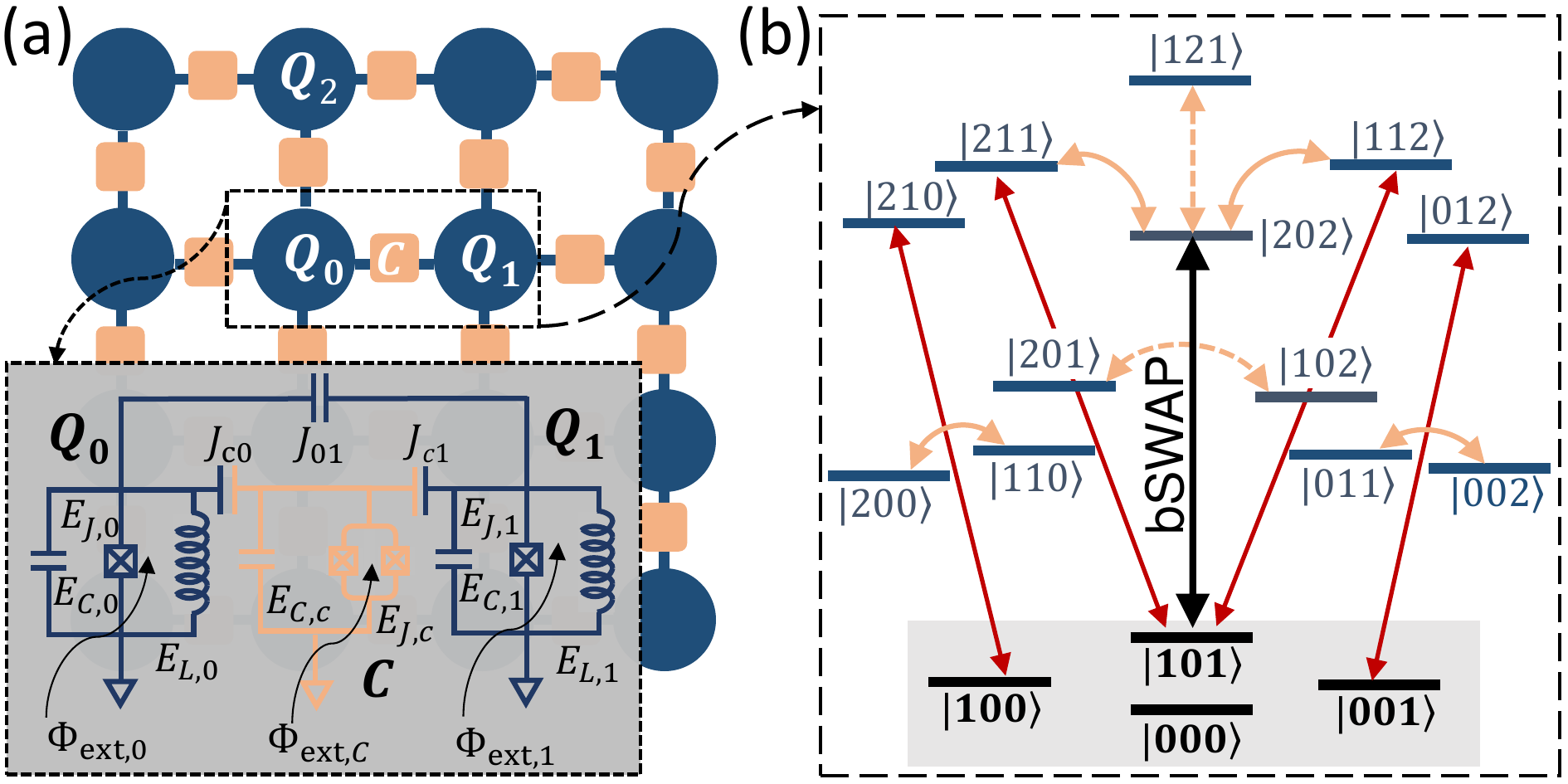}
\end{center}
\caption{(a) A two-dimensional (2D) square qubit lattice comprising fluxoniums (circles) coupled via couplers (squares).
The inset depicts the fluxonium architecture featuring tunable plasmon interactions, where
fluxoniums are coupled via a frequency-tunable transmon coupler. (b) The energy levels of the unit cell
comprising two coupled fluxonium qubits (i.e., $Q_{0}$ and $Q_{1}$), with emphasis on the computational subspace spanned
by $\{|000\rangle, |001\rangle, |100\rangle, |101\rangle\}$ (shaded region) and the fluxonium's plasmon
mode $|1\rangle \rightarrow |2\rangle$. The full system state is labeled as $|Q_0, C, Q_1\rangle$.
Solid orange arrows indicate direct plasmon-coupler couplings, while dashed orange arrows represent coupler-mediated
plasmon-plasmon interactions. In addition to blue sideband transitions between each fluxonium and the
coupler (red arrows), parametric modulation of the coupler can also activate a bSWAP-type
interaction ($|101\rangle \rightarrow |202\rangle$, black arrow) between the plasmon modes
of the two fluxoniums.}
\label{fig1}
\end{figure}

Here we introduce a control strategy that employs parametric modulation of the plasmon interaction to implement fast entangling gates on fluxonium qubits, alternative to the commonly used microwave-based approaches with static plasmon-plasmon couplings~\cite{Nesterov2018,Ficheux2021,Xiong2022,Simakov2023,Ding2023,Rosenfeld2024,Xiong2025,Zhao2025,Singh2025}. Within the fluxonium qubit architecture supporting tunable plasmon interactions~\cite{Zhao2025,Zhao2025b}, fluxoniums are coupled via a tunable coupler, see Fig.~\ref{fig1}(a), as was also very recently demonstrated experimentally in Ref.~\cite{Zhan2026}. The two-qubit gate can be implemented by applying parametric driving to the coupler at the sum frequency of the selective plasmon modes of the two fluxoniums (e.g., $|1\rangle\rightarrow|2\rangle$ for each fluxonium). This drive activates a bSWAP-type interaction~\cite{Bertet2006,Niskanen2007,Roth2017,Poletto2012,Kapit2015,Zhao2017,Nesterov2021} (e.g., $|11\rangle\leftrightarrow|22\rangle$) between the selective plasmon modes, enabling the transition between computational states (e.g, $|11\rangle$) and the non-computational states (e.g, $|22\rangle$), see Fig.~\ref{fig1}(b). Similar to the microwave-based approaches~\cite{Nesterov2018,Simakov2023}, this transition enables the accumulation of a non-trivial conditional phase on the computational state (e.g, $|11\rangle$) and thus facilities the
implementation of controlled-phase gates. We note that, similar to the schemes studied in Refs.~\cite{Moskalenko2021,Moskalenko2022,Didier2018}, employing parametric modulation on the fluxonium qubits could serve as an alternative approach for activating such bSWAP-type transitions.

Considering its features, this strategy offers several potential advantages over traditional
microwave-based methods, particularly in large-scale systems: (i) Driving the coupler directly, rather than the qubits, can potentially reduce control crosstalk and its impact on system performance, an essential feature for densely coupled qubit systems: in a square lattice, see Fig.~\ref{fig1}(a), each qubit can have four nearest neighbors (i.e., four couplers), whereas each coupler has only two nearest neighbors (i.e., two qubits). (ii) The parametric drive frequency can be placed far from frequency bands used for other operations, minimizing unwanted transitions and alleviating spectral crowding. (iii) Parametric-activated gates generally pose less stringent requirements on system parameters (e.g., transition frequencies)~\cite{Roth2017,McKay2016}, offering better resilience to fabrication variations and parameter misalignments.

Despite these advantages, the proposed strategy shares a significant challenge with conventional microwave-based methods, i.e., the short coherence times of plasmon modes (or coupler modes). Furthermore, since our approach involves
doubly excited plasmon states, this coherence limitation is expected to be more severe compared to conventional schemes, which typically utilize only singly excited plasmon states. However, unlike parametric gates in transmon systems, which often exceed 100 ns in gate lengths (see, e.g., Refs.~\cite{McKay2016,Han2020,Li2022,Ganzhorn2020}; a notable exception is the fast parametric scheme in Ref.~\cite{Kubo2023} with gate times below 25 ns), the present
strategy enables sub-100ns CZ gates on fluxoniums with intrinsic errors below $10^{-4}$.
We therefore expect that, even with coherence times on the order of $\sim10\ \mu{\rm s}$, which aligns
with typical values in current devices~\cite{Ficheux2021,Ding2023}, a gate error approaching $10^{-3}$ remains achievable. Moreover, since the decoherence mechanisms of plasmon modes are similar to those of transmon qubits~\cite{Ficheux2021,Nguyen2019}, one can anticipate no fundamental barrier to achieving coherence times approaching $100\ \mu{\rm s}$ in future implementations. Such an improvement could push gate errors toward the $10^{-4}$ level. Given all these considerations, we expect this approach may constitute a promising control strategy for the development of large-scale, high-performance fluxonium-based quantum processors.

This paper is organized as follows. In Sec.~\ref{SecII}, we introduce the model for the fluxonium system with tunable plasmon interactions and derive the effective Hamiltonian of the full system under the parametric modulation.
In Sec.~\ref{SecIII}, we examine various parametric-activated bSWAP-type interactions within the fluxonium system, with a focus on one specific case: the $|11\rangle \leftrightarrow |22\rangle$ transition. In Sec.~\ref{SecIV}, we detail our control strategy for implementing fast, high-fidelity two-qubit controlled-phase (CZ) gates based on the parametric-activated $|11\rangle \leftrightarrow |22\rangle$ transition. Having discussed this gate architecture, we next examine its limitations and opportunities compared with existing schemes, focusing specifically on fluxonium-based approaches in Sec.~\ref{SecV}. In Sec.~\ref{SecVI}, we summarize our main results and discuss potential directions for future work.

\section{The fluxonium system with parametric modulation}\label{SecII}

The fluxonium coupling architecture considered in this work is schematically depicted in Fig.~\ref{fig1}(a).
Within this architecture, the unit cell comprising two coupled fluxonium qubits is modeled by the
Hamiltonian (with $\hbar = 1$ hereafter)
\begin{equation}
\begin{aligned}\label{eq1}
\hat H=& \sum_{k=0,1}[4 E_{C,k} \hat n^2_k + \frac{E_{L,k}}{2}(\hat\varphi_k - \varphi_{\text{ext},k})^2 - E_{J,k}\cos\hat\varphi_k]\\
&+J_{c0}\hat n_0 \hat n_{c}+J_{c1}\hat n_1 \hat n_{c}+J_{01}\hat n_0 \hat n_{1}
\\&+4 E_{C,c} \hat n^2_{c} - E_{J,c}\cos(\frac{\varphi_{\text{ext},c}}{2})\cos\hat\varphi_{c},
\end{aligned}
\end{equation}
where the first and last lines describe the two fluxonium qubits ($k=0,1$) and the frequency-tunable transmon
coupler ($c$), respectively, while the second line accounts for both the fluxonium-coupler couplings and
the direct fluxonium-fluxonium coupling. Here, $E_C$, $E_J$, and $E_L$ denote the charging, Josephson, and
inductive energies, respectively, and $\varphi_\text{ext}$ represents the external phase bias, defined
as $\varphi_\text{ext}=2\pi\Phi_\text{ext}/\Phi_0$ ($\Phi_0$ is the flux quantum). Throughout this work, both fluxonium qubits are biased at
their half-flux-quantum sweet spots with $\varphi_{\text{ext},k}=\pi$, unless stated otherwise.

Owing to the small electric transition dipole moment of the computational (qubit) transition, which effectively
decouples the qubit states from the coupler, we focus on coupler-mediated plasmon-plasmon interactions.
Following Ref.~\cite{Zhao2025}, we approximate the transmon coupler as an anharmonic oscillator and
restrict our attention to one specific plasmon mode per fluxonium (e.g., $|j\rangle \rightarrow |l\rangle$ in $Q_0$ and $|r\rangle \rightarrow |t\rangle$ in $Q_1$). Under these approximations, the system Hamiltonian in Eq.~(\ref{eq1}) can be reduced to the form (see Appendix~\ref{A} for details):
\begin{equation}
\begin{aligned}\label{eq2}
\hat H_{p}=&\sum_{k=0,1}\left[\omega_{p,k}\hat p_{k}^{\dag}\hat p_{k}+g_{p,k}(\hat p_{k}+\hat p_{k}^{\dag})(\hat a_{c}+\hat a_{c}^{\dag})\right]
\\&+\omega_{c}\hat a_{c}^{\dag}\hat a_{c}+\frac{\alpha_{c}}{2}\hat a_{c}^{\dag}\hat a_{c}^{\dag}\hat a_{c}\hat a_{c}
+g_{p,01}(\hat p_{0}+\hat p_{0}^{\dag})(\hat p_{1}+\hat p_{1}^{\dag}),
\end{aligned}
\end{equation}
where
\begin{equation}
\begin{aligned}\label{eq3}
&\hat p_{0}=|j\rangle\langle l|,\,\hat p_{0}^{\dag}=|l\rangle\langle j|,
\\&\hat p_{1}=|r\rangle\langle t|,\,\hat p_{1}^{\dag}=|t\rangle\langle r|,
\end{aligned}
\end{equation}
are the lowering and raising operators for the plasmon modes of the two fluxoniums, with transition
frequencies $\omega_{p,0}$ and $\omega_{p,1}$, respectively, $a_{c}$ ($a_{c}^{\dag}$) denotes the annihilation (creation) operator for the coupler, which has transition frequency $\omega_{c}$ and
anharmonicity $\alpha_{c}$, and $g_{p,k}$ and $g_{p,01}$ represent the coupling strengths of the plasmon-coupler couplings and the direct plasmon-plasmon coupling, respectively.

By considering the system to be in the dispersive regime, where the coupler-plasmon detuning $|\Delta_{p,k}|=|\omega_{p,k}-\omega_{c}|$ is much larger than the coupling strength $g_{p,k}$, an effective Hamiltonian can be derived by eliminating the plasmon-coupler coupling terms in Eq.~(\ref{eq2}) using a Schrieffer-Wolff transformation (SWT)~\cite{Bravyi2011} (see Appendix~\ref{A} for details). Assuming the coupler remains in its ground state, we focus exclusively on the inter-fluxonium
interactions, leading to the following approximate effective Hamiltonian:
\begin{equation}
\begin{aligned}\label{eq4}
\hat H_{p,{\rm eff}}=&\sum_{k=0,1}\left[\omega_{p,k}\hat p_{k}^{\dag}\hat p_{k}\right]
+ g_{p}(\hat p_{0}+\hat p_{0}^{\dag})(\hat p_{1}+\hat p_{1}^{\dag}).
\end{aligned}
\end{equation}
Here, $g_{p}$ denotes the strength of the effective plasmon-plasmon interaction, given by
\begin{equation}
\begin{aligned}\label{eq5}
g_{p}=g_{p,01}+\frac{g_{p,0}g_{p,1}}{2}\left[\sum_{k=0,1}(\frac{1}{\Delta_{p,k}}-\frac{1}{S_{p,k}})\right].
\end{aligned}
\end{equation}
with $S_{p,k}=\omega_{p,k}+\omega_{c}$. Note that for clarity, the frequency renormalization of the plasmon
transitions resulting from plasmon-coupler interactions has been omitted.

Given the tunability of the plasmon interaction through coupler frequency adjustments, as expressed in Eq.~(\ref{eq5}), we now introduce a single-tone parametric drive of the form
\begin{equation}
\begin{aligned}\label{eq6}
\Phi_{{\rm ext},c}(t)=\Phi_{s}+\delta_{\Phi}\cos(\omega_{p}t+\phi_{0})
\end{aligned}
\end{equation}
applied to the coupler to rapidly modulate its frequency~\cite{McKay2016,Roth2017}. Here, $\Phi_{s}$ denotes the static
coupler bias, while $\delta_{\Phi}$, $\omega_{p}$, and $\phi_{0}$ are the amplitude, the frequency, and the initial phase of the drive, respectively. For simplicity, we assume $\phi_{0}=0$ hereafter. Under the small-modulation
condition ($\delta_{\Phi}\ll1$), the coupler frequency under such rapid modulation can be approximated to
first order in $\delta_{\Phi}$ as:
\begin{equation}
\begin{aligned}\label{eq7}
\omega_{c}(\Phi_{{\rm ext},c})\approx\omega_{c}(\Phi_{s})+\frac{\partial \omega_{c}}{\partial \Phi_{{\rm ext},c}}\Big|_{\Phi_{s}}\delta_{\Phi}\cos(\omega_{p}t).
\end{aligned}
\end{equation}

By substituting the expression for $g_{p}$ from Eq.~(\ref{eq5}) with the expanded form of $\omega_{c}(\Phi_{{\rm ext},c})$ and further expanding $g_{p}$ to first order in $\delta_{\Phi}\cos(\omega_{p}t)$~\cite{McKay2016,Roth2017}, we can derive an effective Hamiltonian in a rotating frame defined by the plasmon mode frequencies. Under the assumptions that $\omega_{p,0}>\omega_{p,1}$ and that fast-oscillating terms can be neglected, the effective Hamiltonian takes the form:
\begin{equation}
\begin{aligned}\label{eq8}
\hat H_{p,{\rm eff}}\approx g_{\rm eff}e^{+i\omega_{p}t}\left(e^{-i\Delta_{p,01}t}\hat p_{0}\hat p_{1}^{\dag}+e^{-i S_{p,01}t}\hat p_{0}\hat p_{1}\right)+h.c.
\end{aligned}
\end{equation}
with
\begin{equation}
\begin{aligned}\label{eq9}
g_{\rm eff}&=\delta_{\Phi}\frac{\partial g_{p}}{\partial \Phi_{{\rm ext},c}}\Big|_{\Phi_{s}}
\\&=\delta_{\Phi}\frac{g_{p,0}g_{p,1}}{4}\frac{\partial \omega_{c}}{\partial \Phi_{{\rm ext},c}}\Big|_{\Phi_{s}}\left[\sum_{k=0,1}(\frac{1}{\Delta_{p,k}^{2}}+\frac{1}{S_{p,k}^{2}})\right],
\end{aligned}
\end{equation}
$\Delta_{p,01}=\omega_{p,0}-\omega_{p,1}$, and $S_{p,01}=\omega_{p,0}+\omega_{p,1}$. The first and second
terms in parentheses in Eq.~(\ref{eq8}) describe SWAP-type and bSWAP-type interactions between the plasmon
modes of the two fluxoniums, respectively. It should be noted that, as indicated in Eq.~(\ref{eq4}), although
the coupled fluxonium system inherently exhibits both static SWAP-type and bSWAP-type plasmon interactions, the
plasmon modes of the two fluxoniums are typically far detuned. This detuning renders the contribution from
these static interactions non-dominant under the rotating-wave approximation.

Since the SWAP-type plasmon interaction involves only non-computational states and the present work focuses on
realizing two-qubit gates, we concentrate on the bSWAP-type plasmon interactions (blue sideband transitions), which enable transitions between computational and non-computational plasmon states. As indicated by Eq.~(\ref{eq8}), such interactions are activated when the parametric drive frequency satisfies $\omega_{p}=\omega_{p,0}+\omega_{p,1}$, i.e., matches the sum frequency of the two fluxoniums' plasmon transitions.

Note that while the approximate model derived above offers useful physical insight into parametric-activated plasmon
interactions, it omits several important features. As shown in Fig.~\ref{fig1}(b), in addition to the bSWAP-type interaction, spurious transitions, such as sideband transitions between each fluxonium and the coupler, are also present~\cite{Beaudoin2012} (we will revisit this issue in Sec.~\ref{SecIIIB} and provide further details in Fig.~\ref{fig14} of Appendix~\ref{B}). Furthermore, unlike the expression in Eq.~(\ref{eq5}) derived under the adiabatic approximation, the coupling strengths for SWAP-type and bSWAP-type interactions are generally distinct~\cite{Roth2017}. A more rigorous treatment using time-dependent Schrieffer-Wolff transformation can be found in Refs.~\cite{Roth2017,Petrescu2023}; however, the resulting expressions do not yield significant additional insight beyond the current model. Therefore, we retain the approximate model in the following analysis.

\section{parametric-activated bSWAP-type plasmon interaction}\label{SecIII}

\begin{table}[!htb]
\caption{\label{tab:circuit_parameters} The circuit Hamiltonian parameters of the coupled fluxonium system shown in the
inset of Fig.~\ref{fig1}. The values in parentheses correspond to the configuration with reduced
fluxonium-coupler coupling strength and Josephson energy of the transmon coupler.}
\begin{ruledtabular}
\begin{tabular}{cccc}
(GHz)&
$E_C/2\pi$ &
$E_L/2\pi$ &
$E_J/2\pi$ \\\hline
Fluxonium $Q_{0}$ & 1.41 & 0.80 & 6.27 \\
Fluxonium $Q_{1}$ & 1.30 & 0.59 & 5.71  \\
Transmon $C$ & 0.32 & $-$ & 55 (40) \\
Spectator $Q_{2}$  & 1.33 & 0.60 & 5.40 \\
\hline
\hline
(MHz) & $J_{c0}/2\pi$  & $J_{c1}/2\pi$  & $J_{01}/2\pi$ \\\hline
Coupling strengths & 500 (300) & 500 (300)  & 125 (80)
\end{tabular}
\end{ruledtabular}
\end{table}

\begin{table}[!htb]
\caption{\label{tab:qubit_parameters} The fluxonium frequencies and the maximum coupler frequencies of the coupled fluxonium system with
the circuit Hamiltonian parameters listed in Table~\ref{tab:circuit_parameters}.}
\begin{ruledtabular}
\begin{tabular}{ccccc}
(GHz) &$\omega_{01}/2\pi$ & $\omega_{12}/2\pi$& $\omega_{03}/2\pi$& $\omega_{14}/2\pi$\\\hline
$Q_{0}$ & 0.298 & 5.621 & 8.347 & 12.293\\
$Q_{1}$& 0.222  & 5.269 & 7.461 &  11.019\\
$C$ & 11.537 (9.788) & 11.194 (9.441)  & - & -
\end{tabular}
\end{ruledtabular}
\end{table}

In this section, we present a numerical analysis of bSWAP-type plasmon interactions in the coupled
fluxonium system, based on the approximate model and using the circuit parameters summarized in
Table~\ref{tab:circuit_parameters}. To ensure the feasibility of the proposed gate architecture, we choose parameters that are achievable in current experiments. Accordingly, the fluxonium qubit parameters are adopted from the experimental work in Ref.~\cite{Ding2023}, and the coupling strength we use is well within the experimentally demonstrated value (e.g., $550\,\rm MHz$ in Ref.~\cite{Ding2023}). The numerical analysis presented below is performed by solving the Schr\"{o}dinger equation with the system Hamiltonian given in Eq.~(\ref{eq1}) and the parametric drive given in Eq.~(\ref{eq6}), and is carried out using the Python package QuTiP~\cite{Johansson2012}. For convenience, the corresponding fluxonium and coupler frequencies are provided in Table~\ref{tab:qubit_parameters}. Prior to detailing
the system's behavior under parametric driving, we first examine the tunable plasmon interactions
within the architecture. Hereafter, the system state is denoted as $|Q_1,C,Q_2\rangle$ and when referring
exclusively to the fluxonium subspace, the notation $|Q_1,Q_2\rangle\equiv|Q_1,0,Q_2\rangle$ is used.

\subsection{Tunable plasmon interaction}\label{SecIIIA}

\begin{figure}[tbp]
\begin{center}
\includegraphics[keepaspectratio=true,width=\columnwidth]{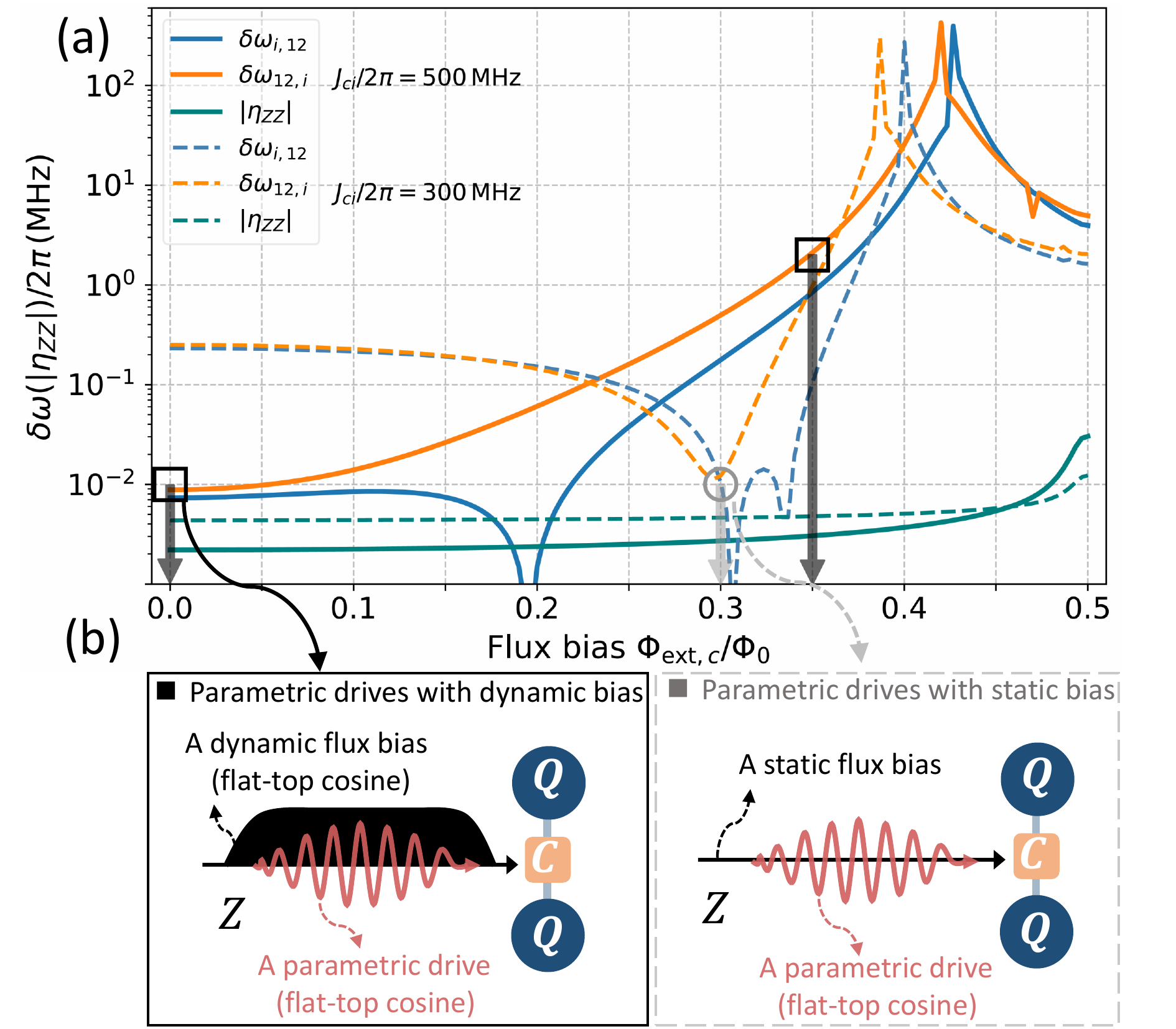}
\end{center}
\caption{(a) Coupler-mediated interactions for the plasmon transition $|1\rangle\rightarrow|2\rangle$, characterized by
state-dependent plasmon frequency shifts as a function of coupler flux bias. Discontinuities and abrupt jumps in the
curves result from state labeling ambiguities near avoided crossings. The ZZ coupling $\eta_{ZZ}$ is also shown and typically remains below $10\,\rm kHz$. Solid and dashed lines represent results for
coupled fluxonium systems with distinct parameter sets, including coupling strengths and coupler frequencies, as
specified in Table~\ref{tab:circuit_parameters}. Black arrows at biases of 0.0 and 0.35 indicate the coupler idle point (where state-dependent frequency shifts are minimized) and interaction point, respectively, for the system with $J_{ck}/2\pi=500\,{\rm MHz}$, while the gray arrow at a bias of 0.3 marks both the idle and the interaction points for the system with $J_{ck}/2\pi=300\,{\rm MHz}$. (b) In the context of implementing parametric gates, these
parameter sets lead to two distinct operational configurations: the left panel illustrates the combination of a
parametric drive with a dynamic flux bias, while the right panel shows the combination of a parametric drive
with a static flux bias.}
\label{fig2}
\end{figure}

\begin{figure*}[htbp]
\begin{center}
\includegraphics[width=16cm,height=6cm]{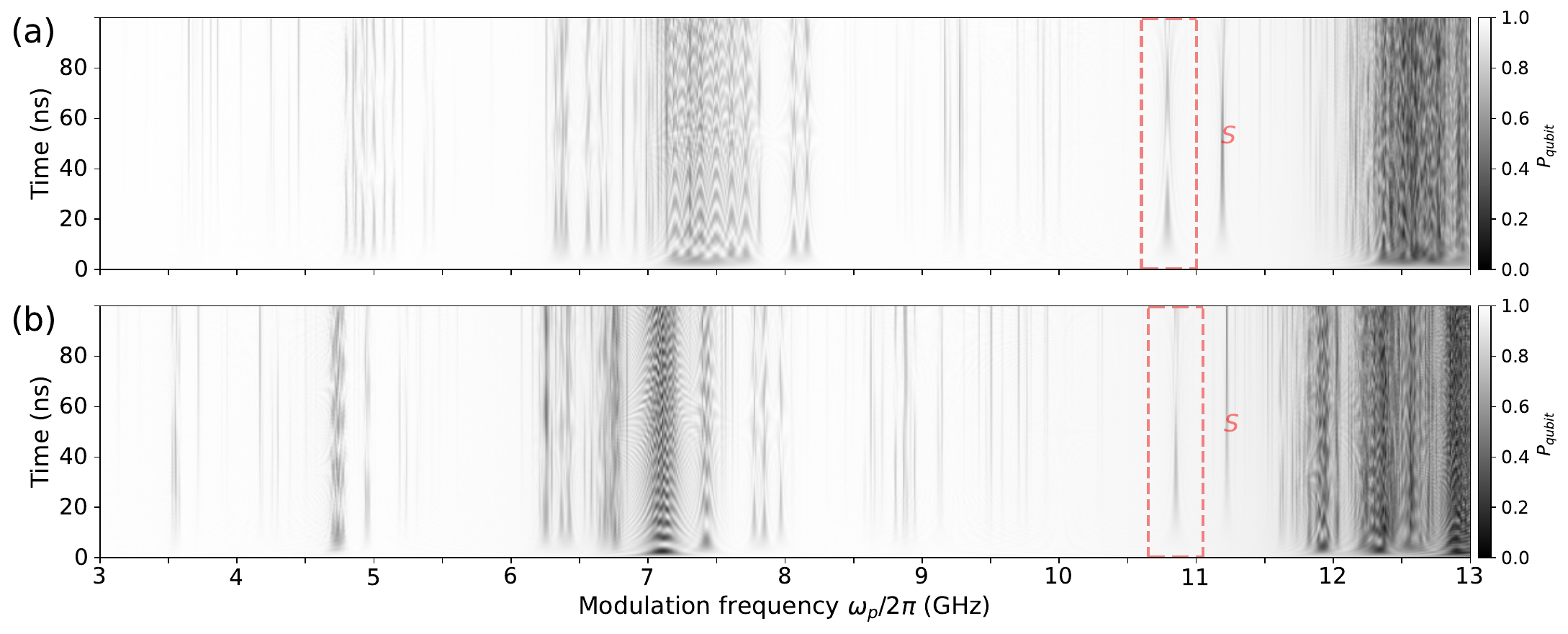}
\end{center}
\caption{Population within the computational (qubit) subspace, i.e., $P_{qubit}=P_{00}+P_{01}+P_{10}+P_{11}$, as a function of parametric drive frequency and
evolution time, for the coupled fluxonium system initialized in the state $(|00\rangle+|01\rangle+|10\rangle+|11\rangle)/2$.
(a) System with $J_{ck}/2\pi=500\,{\rm MHz}$. The static coupler bias is set to $\Phi_{s}/\Phi_{0}=0.35$,
and the parametric drive amplitude is $\delta_{\Phi}/\Phi_{0}=0.045$. (b) System with $J_{ck}/2\pi=300\,{\rm MHz}$.
The static coupler bias is set to $\Phi_{s}/\Phi_{0}=0.30$, and the drive amplitude is
$\delta_{\Phi}/\Phi_{0}=0.075$. The pink dashed boxes highlight the parametric-activated transition
$|11\rangle\rightarrow|22\rangle$, while the pink label $S$ indicates an example of a spurious transition induced by
the parametric drive, specifically $|000(100)\rangle\leftrightarrow|004(104)\rangle$; see Fig.~\ref{fig13} in Appendix~\ref{B} for a detailed illustration of these spurious transitions.}
\label{fig3}
\end{figure*}

For illustration purposes only, here we focus on the coupled-mediated interaction for the plasmon transition
$|1\rangle\rightarrow|2\rangle$. As demonstrated in previous studies~\cite{Nesterov2018,Zhao2025}, the inter-fluxonium
plasmon interaction can induce conditional frequency shifts in the plasmon transition. Accordingly, to quantify the
interaction strength, we adopt the metric of state-dependent frequency shifts~\cite{Zhao2025}, defined as
\begin{equation}
\begin{aligned}\label{eq10}
&\delta\omega_{p,0}=|(E_{21}-E_{11})-(E_{20}-E_{10})|,\\
&\delta\omega_{p,1}=|(E_{12}-E_{11})-(E_{02}-E_{01})|,
\end{aligned}
\end{equation}
for the two fluxoniums, respectively, where $E_{kl}$ denotes the energy of the system eigenstate $|kl\rangle$. Using
the circuit parameters in Table~\ref{tab:circuit_parameters}, Figure~\ref{fig2}(a) shows the plasmon interaction-induced shift as a function of coupler flux bias for two distinct parameter sets. 
The corresponding ZZ coupling strength, $\eta_{ZZ}=(E_{11}-E_{01})-(E_{10}-E_{00})$, is also shown and typically remains below $10\,\rm kHz$, confirming that the computational states are effectively decoupled.

As indicated by Eq.~(\ref{eq9}), the strength of the parametric-activated interaction is proportional to the derivative
of the plasmon interaction with respect to the coupler flux bias, i.e., $\partial g_{p}/\partial \Phi_{{\rm ext},c}$.
Thus, as shown in Figs.~\ref{fig2}(a) and~\ref{fig2}(b), two operational configurations can be identified for
implementing parametric gates. When the derivative at the system (coupler) idle
point, where state-dependent frequency shifts are minimized, is insufficient to achieve strong parametric-activated
interaction, a dynamic bias is required to shift the coupler from its idle point to an interaction point that provides
a larger derivative, as exemplified by systems with $J_{ck}/2\pi=500\,{\rm MHz}$. Conversely, when the derivative at the idle point already supports strong parametric-activated interaction, only a static bias is needed, which is the case for systems with $J_{ck}/2\pi=300\,{\rm MHz}$.

Note that from a control complexity perspective, the static-bias configuration is often preferable~\cite{Li2022}. However, successful implementation of parametric gates under this approach may impose stricter constraints on circuit parameters. In contrast, the dynamic flux-bias configuration can offer greater flexibility in selecting bias parameters~\cite{Han2020} and provide operational redundancy to mitigate challenges such as spurious couplings to defect modes~\cite{Muller2019,Klimov2018,Wang2023,Wang2025a,Chen2025}.

\subsection{Parametric-activated interaction}\label{SecIIIB}

Here we turn to analyze parametric-activated interactions in the coupled fluxonium system, with particular
emphasis on the bSWAP-type interaction of $|11\rangle\leftrightarrow|22\rangle$. Before examining this
specific interaction in detail, we provide a broader overview of the parametric processes activated by rapid
flux modulation of the coupler frequency. This overview could offer insight beyond the approximate model
introduced in Sec.~\ref{SecII} and helps illustrate both potential challenges and opportunities for
implementing fast, high-fidelity entangling gates in this architecture.

Under the parametric drive described by Eq.~(\ref{eq6}), Figure~\ref{fig3} displays the population within
the computational subspace ($P_{qubit}=P_{00}+P_{01}+P_{10}+P_{11}$) as a function of modulation frequency and evolution time for the coupled fluxonium
system initialized in the state $(|00\rangle+|01\rangle+|10\rangle+|11\rangle)/2$. In Fig.~\ref{fig3}(a), with a
static coupler bias of $\Phi_{s}/\Phi_{0}=0.35$, a modulation amplitude of $\delta_{\Phi}/\Phi_{0}=0.045$, and a
coupling strength of $J_{ck}/2\pi=500\,{\rm MHz}$, numerous chevron patterns are observed across the frequency range
from 3 GHz to 13 GHz. Each pattern generally corresponds to a specific parametric-activated transition. Similar behavior is evident in Fig.~\ref{fig3}(b) for the system with $J_{ck}/2\pi=300\,{\rm MHz}$, where the static coupler bias and drive amplitude are $\Phi_{s}/\Phi_{0}=0.30$ and $\delta_{\Phi}/\Phi_{0}=0.075$, respectively.

As previously noted, the intrinsic nonlinearity of the transmon coupler enables rapid flux modulation
of its frequency to activate both bSWAP-type transitions between coupled plasmon modes and various
other transitions, such as blue sideband transitions between fluxonium plasmon modes and the
coupler (see Appendix~\ref{B1} for details). Based on their physical origins, these transitions can be
classified into three main categories:

\textbf{(1) bSWAP-type transitions for plasmon mode pairs}, including transitions within the same plasmon mode
of the two fluxoniums (e.g., $|11\rangle\rightarrow|22\rangle$, highlighted by the pink dashed boxes,
and $|00\rangle\rightarrow|33\rangle$) and between different modes (e.g., $|01\rangle\rightarrow|34\rangle$
and $|10\rangle\rightarrow|23\rangle$), as suggested by the discussion given in Sec.~\ref{SecII};

\textbf{(2) Blue sideband transitions between the fluxonium plasmon modes and the coupler}, such
as $|001\rangle\rightarrow|014\rangle$ and $|001\rangle\rightarrow|311\rangle$;

\textbf{(3) Coupler state excitations due to effective two-photon (squeezing) drives}, for
example $|000\rangle\rightarrow|020\rangle$, which arise from the flux modulation of the coupler's nonlinear
potential~\cite{Didier2018}, see Appendix~\ref{B2} for details.

In addition to these main categories, strong state hybridization among fluxonium plasmon modes and the coupler, resulting from their strong couplings, can induce parametric-activated cross-driving transitions (facilitated by the three main types of transitions described above). These are analogous to cross-resonance
effects in microwave-driven coupled qubit systems~\cite{Paraoanu2006,Rigetti2010,de Groot2010,Chow2011} and can give rise to various high-energy transitions through single- or multi-photon processes when on-resonance conditions (frequency collisions) are inadvertently met~\cite{Malekakhlagh2020,Heya2024,Zhao2024,Malekakhlagh2022}. A typical example is the transition $|000\rangle\rightarrow|004\rangle$ (labeled $S$, see Fig.~\ref{fig3}), which is prohibited
for the bare fluxonium at the half flux quantum sweet spot but becomes allowed in the coupled system due to
hybridization between between $|004\rangle$ and $|013\rangle$.

\begin{figure}[tbp]
\begin{center}
\includegraphics[keepaspectratio=true,width=\columnwidth]{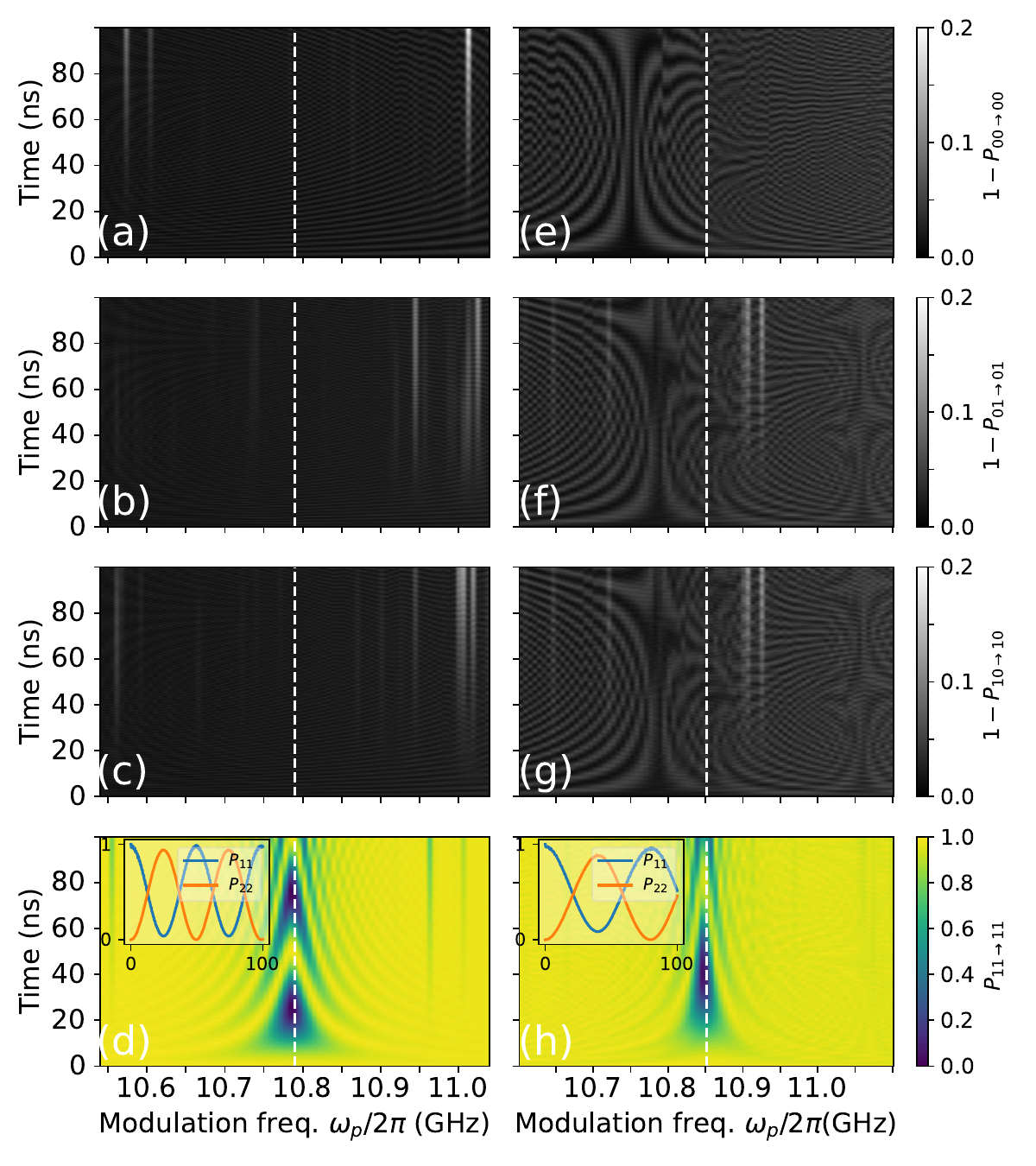}
\end{center}
\caption{Population versus the parametric drive frequency around the $|11\rangle\rightarrow|22\rangle$ transition and
the evolution time, presented as an enlarged view of the region within the pink dashed box in Fig.~\ref{fig3}.
Here, $P_{ij\rightarrow ij}$ represents the population in state $|ij\rangle$ when the system is initially prepared
in $|ij\rangle$. The circuit parameters used in (a-d) and (e-h) are the same as those in Fig.~\ref{fig3}(a) and
Fig.~\ref{fig3}(b), respectively. The vertical dashed lines indicate the ideal transition frequency for $|11\rangle\rightarrow|22\rangle$ without accounting for drive-induced frequency shifts, while the corresponding system dynamics (populations in the $|11\rangle$ and $|22\rangle$ states) with the system prepared in $|11\rangle$ are shown in the insets of (d) and (h). Note that the faint oscillations observed here are generally attributed to higher-order spurious transitions.}
\label{fig4}
\end{figure}

\begin{table}[!htb]
\caption{\label{tab:transition_parameters} Transition parameters at the interaction point for the coupled fluxonium system with
the circuit Hamiltonian parameters listed in Table~\ref{tab:circuit_parameters}.}
\begin{ruledtabular}
\begin{tabular}{ccccc}
$ $& $|0\langle|\hat{n}_{k}|1\rangle|$& $ |1\langle|\hat{n}_{k}|2\rangle|$& $|0\langle|\hat{n}_{k}|3\rangle|$ & $|1\langle|\hat{n}_{k}|4\rangle|$\\\hline
$Q_{0}$ & 0.068 & 0.562 & 0.488 & 0.214\\
$Q_{1}$& 0.057  & 0.557 & 0.498 &  0.202\\
\hline
\hline
C ($\Phi_{s}$) & $\omega_{01}/2\pi$  & $\omega_{12}/2\pi$  & $|0\langle|\hat{n}_{c}|1\rangle|$ & $ |1\langle|\hat{n}_{c}|2\rangle|$ \\\hline
0.35  & 7.661 & 7.305   & 1.223  & 1.689\\
0.30 & 7.423 & 7.066  & 1.204 & 1.661\\
\end{tabular}
\end{ruledtabular}
\end{table}

\begin{figure}[tbp]
\begin{center}
\includegraphics[keepaspectratio=true,width=\columnwidth]{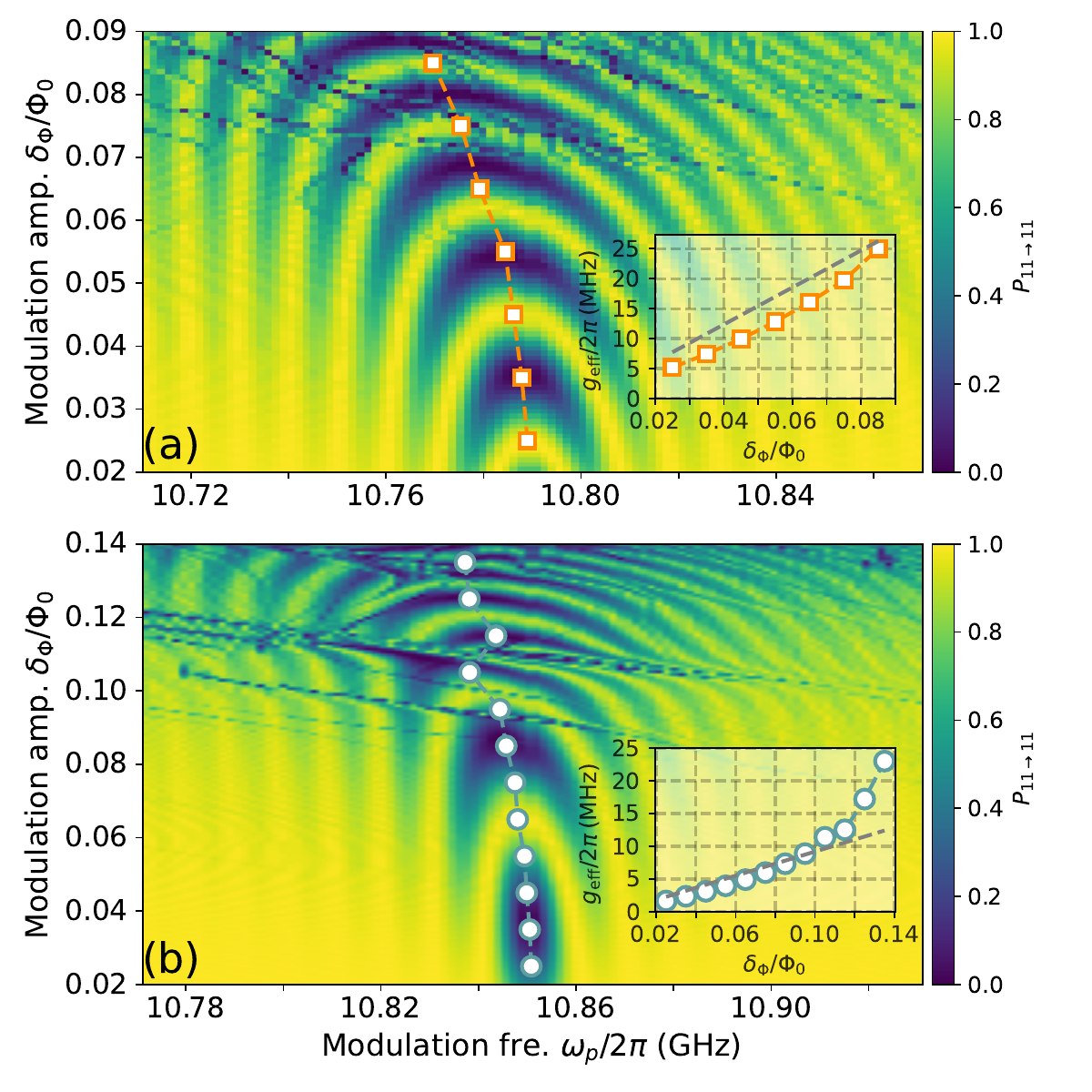}
\end{center}
\caption{Population in $|11\rangle$ in the parametric-driven system as a function of the parametric drive
frequency around the $|11\rangle\rightarrow|22\rangle$ transition and the drive amplitude, with the evolution
time fixed at 100 ns and the initial state of $|11\rangle$. The circuit parameters used in (a) and (b) correspond
to those in Fig.~\ref{fig3}(a) and Fig.~\ref{fig3}(b), respectively. The orange square in (a) and the teal circle
in (b) mark the transition frequencies of $|11\rangle\rightarrow|22\rangle$ obtained using the Floquet numerical
method. The corresponding transition strengths are shown in the insets, with grey lines indicating the
results from the approximate model.}
\label{fig5}
\end{figure}

In general, aside from the targeted transitions such as $|11\rangle\rightarrow|22\rangle$ (see Fig.~\ref{fig4} for an
enlarged view of the area within the dashed pink box in Fig.~\ref{fig3}) studied here for
realizing two-qubit gates, all other transitions, particularly those spectrally close to the target transition, can
complicate gate control and degrade performance. Furthermore, when seeking higher gate speeds through stronger
parametric drives or coupling strengths, higher-order spurious transitions, though typically faint here (see Figs.~\ref{fig3} and~\ref{fig4}), may become non-negligible. These unwanted transitions introduce frequency collision issues that intensify with increasing drive magnitudes and coupling strengths.

However, unlike transmon-based systems~\cite{Osman2023}, the fluxonium qubit benefits from its strong anharmonicity, small qubit transition dipole moment, and transmon-like plasmon transition dipoles. This enables the implementation
of strong parametric-activated interactions without significant interference from other spurious transitions, even
when the coupled plasmon-coupler system operates in the non-dispersive regime~\cite{Goerz2017,Sung2021,Zhao2021} (see the coupling parameters listed in Table~\ref{tab:transition_parameters}) and under large modulation amplitudes. Specifically, as shown in Figs.~\ref{fig3} and~\ref{fig4}, the target $|11\rangle\rightarrow|22\rangle$ transition remains well separated from other significant spurious transitions and there does not exhibit notable frequency crowding issues, even at activated strengths exceeding 5 MHz (corresponding to oscillation periods below 100 ns).

To explore the limits of such collision-free behavior, Figure~\ref{fig5} shows the population in $|11\rangle$ under flux modulation in the coupled system, which is initially prepared in $|11\rangle$. The population is plotted as a function of both the parametric drive frequency around the $|11\rangle\rightarrow|22\rangle$ transition and the drive amplitude, with the evolution time fixed at 100 ns.  In addition, we also employ the Floquet numerical method~\cite{Petrescu2023}, specifically by numerically computing the quasienergy spectrum of the periodically driven Hamiltonian, also referred to as the Floquet Hamiltonian~\cite{Shirley1965,Sambe1973,Johansson2012}, to extract the transition frequencies and strengths (see the inset with grey lines indicating the
results from Eq.~(\ref{eq9}) of the approximate model). It can be observed that even when further increasing the modulation strength, no significant spurious transitions emerge as the strength of the parametric-activated bSWAP interaction reaches approximately $15\,\rm MHz$ and $10\,\rm MHz$ (substantially larger than those achieved in transmon-based systems~\cite{Roth2017,McKay2016,Han2020,Li2022,Ganzhorn2020}) for systems with
coupling strengths $J_{ck}/2\pi=500\,{\rm MHz}$ and $J_{ck}/2\pi=300\,{\rm MHz}$, respectively. These results
suggest that entangling gates based on the parametric-activated $|11\rangle\rightarrow|22\rangle$ transition
can be successfully implemented with gate lengths below 100 ns, as will be illustrated in the following
section. Furthermore, we note that the leftward tilt of the pattern in Fig.~\ref{fig5} (i.e., the optimal modulation frequncy $\omega_{p}$ for activating the $|11\rangle\leftrightarrow|11\rangle$ transition) is caused by the parametric-drive-induced frequency shift (shift down) of the fluxonium transition $|1\rangle\leftrightarrow|2\rangle$, as illustrated in Ref.~\cite{Roth2017}.

\section{CZ gate implementations based on the bSWAP interaction}\label{SecIV}

Given the availability of strong parametric-activated bSWAP interactions in the current architecture, here
we turn to employ them to implement fast entangling gates. While the following discussion focuses on the bSWAP
interaction between the plasmon modes $|1\rangle\rightarrow|2\rangle$ of the two fluxonium qubits, we note
that other types of activated bSWAP interactions, e.g., $|00\rangle\rightarrow|33\rangle$
and $|10\rangle\rightarrow|23\rangle$ as discussed in Sec.~\ref{SecIIIB}, can also be utilized
for the same purpose. However, beyond the circuit Hamiltonian parameters utilized here to enhance
the bSWAP-type interaction $|11\rangle\rightarrow|22\rangle$, the implementation of other such interactions
requires careful optimization of circuit parameters, especially plasmon mode frequencies, to achieve large
activated coupling strengths and reduced frequency collisions.

\begin{figure}[tbp]
\begin{center}
\includegraphics[keepaspectratio=true,width=\columnwidth]{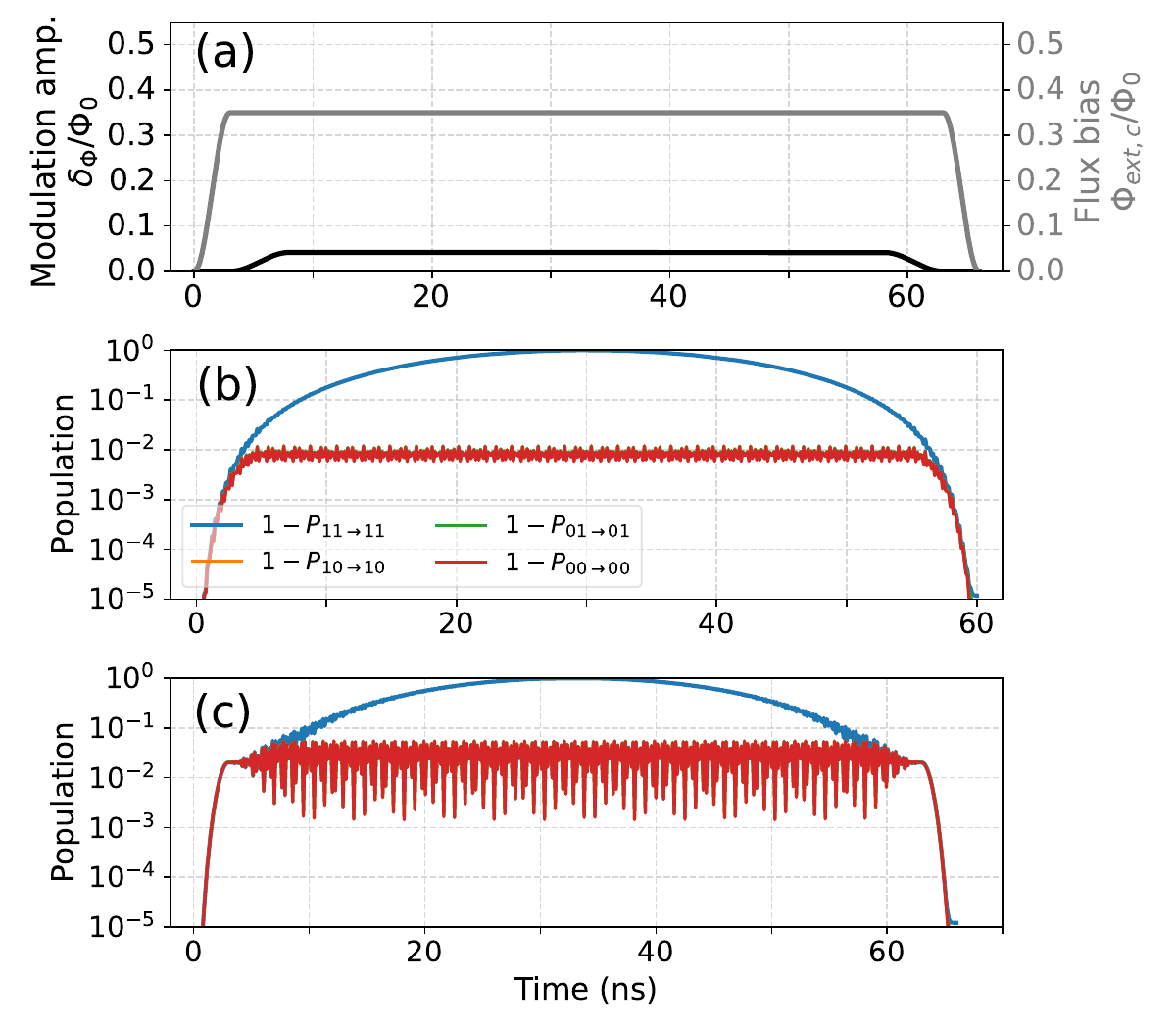}
\end{center}
\caption{(a) A typical control pulse (flat-top cosine) for implementing CZ gates in the parametric-driven fluxonium
system, comprising a dynamic flux bias pulse (biasing the coupler from its idle point to the interaction point, as shown in Fig.~\ref{fig2}) and the envelope of the parametric drive pulse. The flux pulse has a ramp time of 3 ns, and the drive pulse ramp time is 5 ns. (b) and (c) show typical system dynamics during the parametric-activated CZ gate operation without and with the dynamic flux bias, respectively. The used circuit Hamiltonian parameters are the same
as those in Fig.~\ref{fig3}(a). Here, $P_{ij\rightarrow ij}$ denotes the population remaining in
state $|ij\rangle$ after initializing the system in $|ij\rangle$.}
\label{fig6}
\end{figure}

\begin{figure}[tbp]
\begin{center}
\includegraphics[keepaspectratio=true,width=\columnwidth]{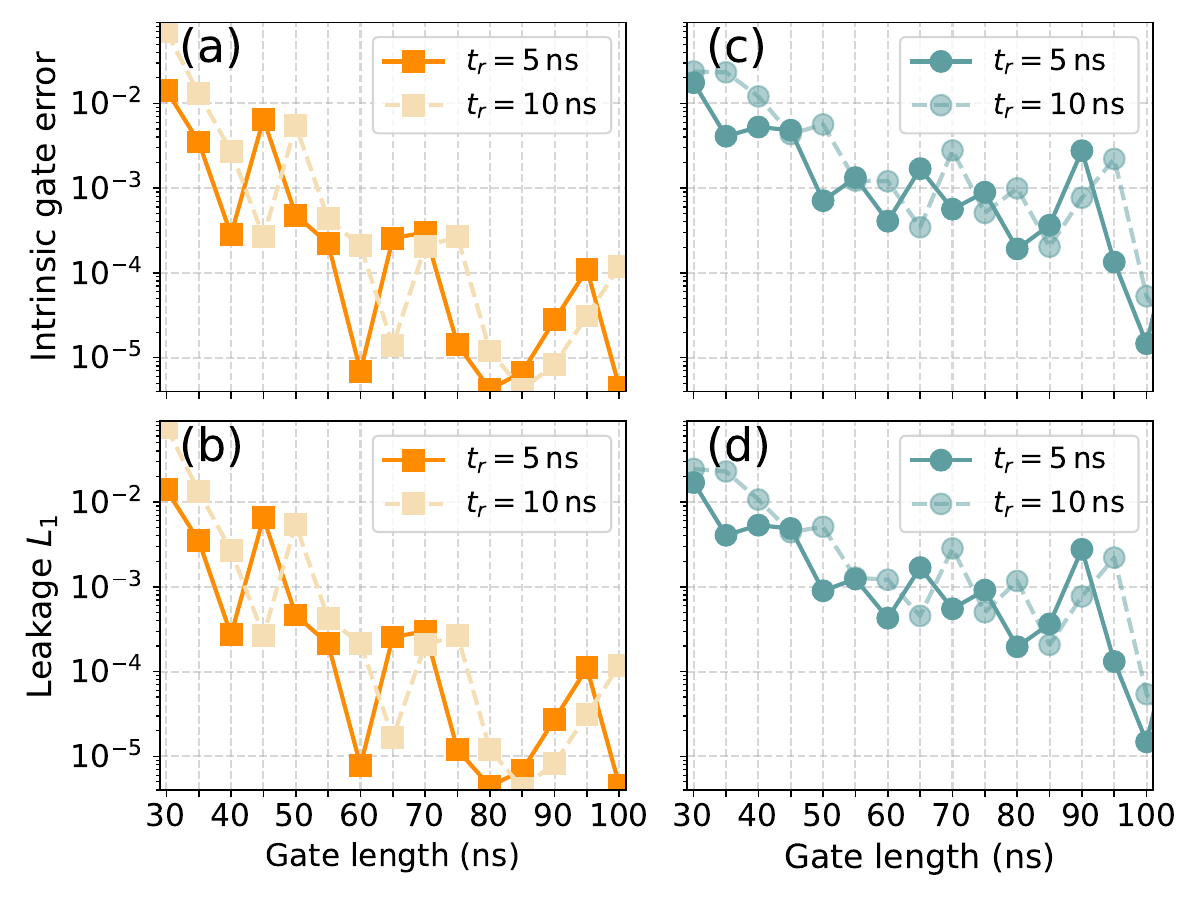}
\end{center}
\caption{(a, c) Intrinsic gate errors (excluding qubit and coupler relaxation and dephasing) and (b, d) leakage
of the parametric CZ gates as functions of gate
length. (a, b) and (c, d) correspond to the dynamic flux-bias configuration and the static bias
configuration, respectively, see Fig.~\ref{fig2}(b). Solid and dashed lines represent results for drive pulse ramp times
of 5 ns and 10 ns, respectively.}
\label{fig7}
\end{figure}

\subsection{Gate error}\label{SecIVA}

As shown in Fig.~\ref{fig2}(b), we consider two operational configurations: the dynamic flux-bias configuration
and the static bias configuration. Figure~\ref{fig6}(a) displays a typical control pulse (flat-top
cosine pulses, see Appendix~\ref{C} for details) used to engineer the bSWAP-type interaction $|11\rangle\rightarrow|22\rangle$ for implementing CZ gates. Note that the adoption of finite ramp times for the control pulse serves both to accommodate control electronics constraints in practical systems and to suppress potential leakage from off-resonance transitions~\cite{Barends2019,Malekakhlagh2022b}. In the dynamic flux-bias configuration, the control pulse consists of a dynamic flux bias pulse, which tunes the coupler from its idle point (i.e., $\Phi_{{\rm ext},s}/\Phi_{0}=0$) to the interaction point (i.e.,$\Phi_{{\rm ext},s}/\Phi_{0}=0.35$), and a parametric drive pulse. In contrast, the static bias configuration requires only a parametric drive pulse (also adopting a flat-top cosine shape, similar to the pulse used for the dynamic flux-bias configuration shown in Fig.~\ref{fig6}(a)), with the coupler remaining at its idle point (i.e., $\Phi_{{\rm ext},s}/\Phi_{0}=0.30$). Figures~\ref{fig6}(b) and~\ref{fig6}(c) illustrate the corresponding gate dynamics for the dynamic flux-bias configuration without and with the dynamic flux bias applied, respectively. The population distributions at the end of the gate operation are nearly identical in both cases, further confirming that the qubit states are effectively decoupled within this architecture and that non-adiabatic transitions during coupler bias ramping are negligible~\cite{Zhao2025}.

To characterize the intrinsic gate performance (excluding qubit and coupler relaxation and
dephasing), we consider the metric of state-average gate fidelity~\cite{Pedersen2007}, see Appendix~\ref{C} for detail. For each gate length and the coupler bias setting, optimal drive parameters, including the parametric drive frequency and amplitude, can be obtained by minimizing both leakage~\cite{Wood2018} and conditional phase error~\cite{Zhao2025,Jiang2025} (see Appendix~\ref{C} for detail). As shown in Figs.~\ref{fig7}(a) and~\ref{fig7}(b), the resulting gate error and leakage are shown as functions of gate length for both operational configurations. These results reveal a clear speed-fidelity trade-off: longer gate durations enable lower gate errors rates, and the gate performance is mainly limited by leakage into noncomputational states.

Besides their overall decreasing trends, both gate error and leakage also exhibit identical oscillatory behavior
as a function of gate length. These features are commonly attributed to spurious off-resonant
interactions~\cite{Zhao2025}, as will be discussed in the following subsection. Furthermore, varying the
drive ramp times, such as increasing the ramp time from 5 ns to 10 ns, while keeping the
length of the pulse flat-top constant, does not significantly alter the gate error or leakage of parametric
gates with identical flat-top durations. Similar behavior is observed for both operational configurations, even
though the bSWAP-type interactions are activated at markedly different drive frequencies. These results suggest
that the dominant leakage mechanism may be unrelated to parametric-activated spurious transitions and is therefore
independent of the drive frequency.

\begin{figure}[htbp]
\begin{center}
\includegraphics[keepaspectratio=true,width=\columnwidth]{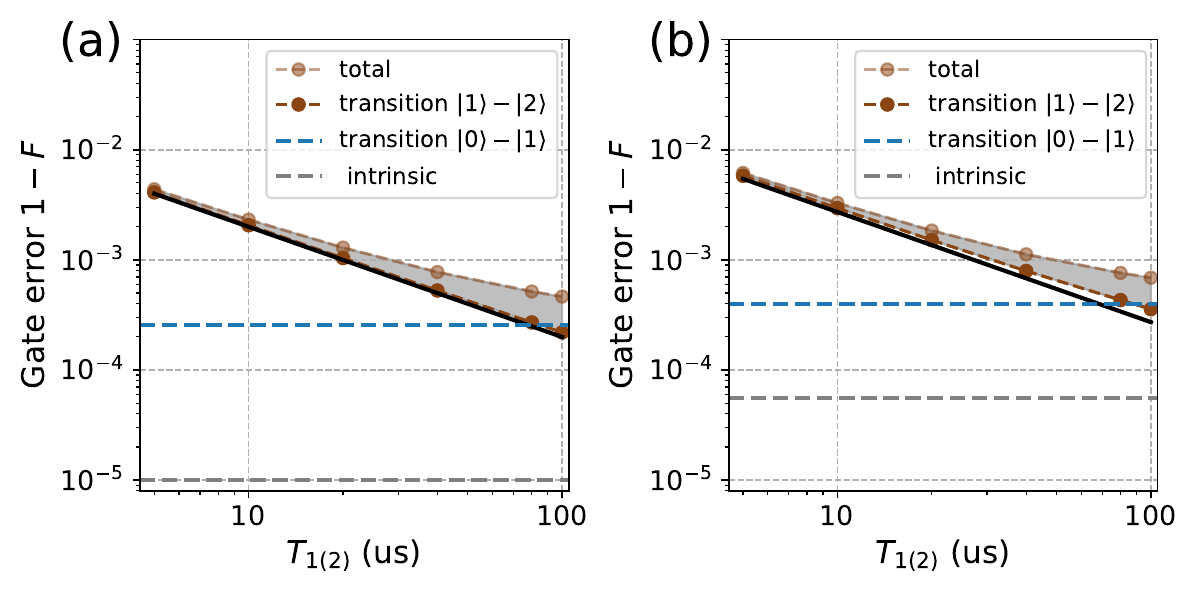}
\end{center}
\caption{Total gate error, gate error assuming only relaxation and dephasing of the plasmon
transition $|1\rangle-|2\rangle$, and gate error assuming only relaxation and dephasing of qubit
transition $|0\rangle-|1\rangle$ as function of the relaxation and dephasing times. For easy
reference, the intrinsic gate errors are also indicated. Here, we assume $T_{1}=1\,\rm ms$ and $T_{\phi}=0.2\,\rm ms$ for the fluxonium computational states $|0\rangle-|1\rangle$, and set $T_{1}=T_{2}$ (with $1/T_2=1/(2T_1)+1/T_{\phi}$) for fluxonium plasmon transitions ($|1\rangle-|2\rangle$, $|0\rangle-|3\rangle$, $|1\rangle-|4\rangle$) and the
coupler. Additionally, the relaxation time of fluxonium plasmon modes is taken as half that
of the transmon coupler. For the leading incoherent error, i.e., that arising from relaxation and dephasing of the plasmon transition $|1\rangle-|2\rangle$, the black solid lines represent the analytical results, which agree well with the numerical simulations. (a) and (b) correspond to the 55-ns CZ gate
in dynamic flux-bias configuration and the 75-ns CZ gate in static bias
configuration, respectively. Here, the drive pulse ramp time is 5 ns.}
\label{fig8}
\end{figure}

Moreover, as shown in Fig.~\ref{fig8}, we assess the impact of fluxonium and coupler relaxation and
dephasing on the proposed gate implementation. Without loss of generality, we consider 55-ns and 75-ns 
CZ gates in the two operational configurations, each with a drive pulse ramp time of 5 ns. We assume the computational states ($|0\rangle-|1\rangle$) of the fluxonium have relaxation and dephasing times of $T_{1}=1\,\rm ms$ and $T_{\phi}=0.2\,\rm ms$ (white noise), consistent with recent experiments (Ref.~\cite{Ding2023}). For the fluxonium plasmon transitions (e.g., $|1\rangle-|2\rangle$, $|0\rangle-|3\rangle$, $|1\rangle-|4\rangle$) and the coupler transitions, we set $T_{1}=T_{2}$ (note here $T_{2}$ denotes the qubit coherence time, with $1/T_2=1/(2T_1)+1/T_{\phi}$). Reflecting current technological capabilities, we further assume the fluxonium plasmon modes have a relaxation time half that of the transmon coupler. Using these parameters and solving the Lindblad master
equation (see Appendix~\ref{C1} for details)~\cite{Wood2018,Zhao2022c}, Figure~\ref{fig8} presents the total gate error, including all decoherence channels, as a function of the relaxation and dephasing times. For comparison, the intrinsic gate errors are also indicated for both configurations. It can be seen that, with typical coherence times on the order of $\sim10\ \mu{\rm s}$ for the plasmon modes (consistent with values reported in current devices~\cite{Ficheux2021,Ding2023}), fast entangling gates can achieve errors approaching $10^{-3}$, which is comparable to state-of-the-art experimental demonstrations~\cite{Ding2023,Zhang2024,Lin2025}. We note that the gate error here is dominated by incoherent errors; consequently, the overall performance can be further enhanced by shortening the gate length (e.g., to 40 ns and 50 ns in the two operational configurations), as illustrated
in Fig.~\ref{fig7}. Thus, we expect that even when accounting for these decoherence processes, a gate error
below $10^{-3}$ should be achievable here.

To further identify the dominant error source, we also plot the errors under two simplified decoherence models: one including only relaxation and dephasing of the plasmon transition $|1\rangle-|2\rangle$ that participates in the
activated bSWAP-type interaction ($|11\rangle\leftrightarrow|22\rangle$), and another including only
decoherence of the computational states ($|0\rangle-|1\rangle$). As expected, for high-coherence fluxoniums, the
leading gate error arises from relaxation and dephasing of the non-computational gate transition
$|1\rangle\leftrightarrow|2\rangle$, consistent with earlier
studies~\cite{Ficheux2021,Ding2023,Abad2023,Zhao2025}. In addition, for the leading incoherent error channel, i.e., relaxation and dephasing of the plasmon transition $|1\rangle-|2\rangle$, we also present the analytical results given in Appendix~\ref{C2}, which agree well with the numerical simulations (see black solid lines).

\subsection{Leakage error analysis}\label{SecIVB}

\begin{figure}[htbp]
\begin{center}
\includegraphics[keepaspectratio=true,width=\columnwidth]{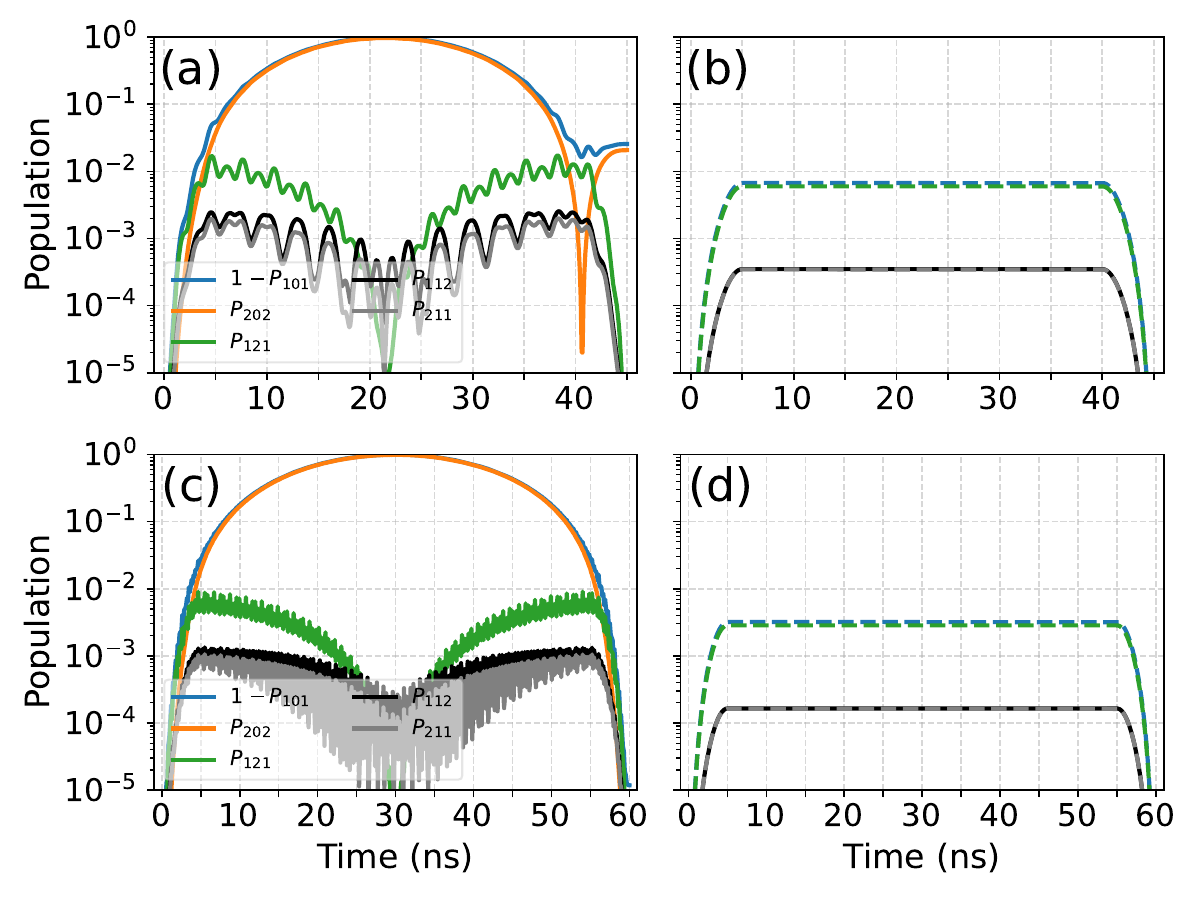}
\end{center}
\caption{Typical system dynamics of CZ gates with and without significant leakage. (a) Dynamics during a
CZ gate of length 45 ns and pulse ramp time 5 ns, applied to the system initialized in $|101\rangle$.
(b) Dynamics under identical parameters to (a), but with the drive frequency set to zero. (c, d) show
results corresponding to (a) and (b), respectively, for a gate length of 65 ns and pulse ramp time of 5 ns.}
\label{fig9}
\end{figure}

As shown in Fig.~\ref{fig8}, the dominant error source in current devices is expected to be relaxation
and dephasing of noncomputational states, based on existing demonstrated technology~\cite{Ficheux2021,Ding2023}. These errors can be mitigated in future implementations through improved fabrication techniques and optimization of qubit
parameters and layout. However, although leakage may not represent the dominant error source in this
gate architecture, it must still be carefully evaluated and minimized, particularly in the context
of quantum error correction~\cite{Miao2023}.

To identify the dominant leakage channel, we examine the system dynamics of CZ gates both with and without
significant leakage. Figure~\ref{fig9}(a) displays the system dynamics for a gate with substantial leakage
at a gate length of 45 ns and a pulse ramp time of 5 ns. The dominant leakage channel is observed to be
from $|101\rangle$ to $|121\rangle$. To verify the conjecture discussed in the previous subsection, Figure~\ref{fig9}(b) shows the system dynamics under the same parameters as in Fig.~\ref{fig9}(a), but with the drive frequency set to zero, i.e., effectively replacing the parametric drive with a dynamic flux pulse of identical
envelope. It can be found that although no significant leakage remains at the end of the pulse, noncomputational
states such as $|211\rangle$, $|112\rangle$, and $|121\rangle$ may still be temporarily populated during the
flux pulse. As anticipated in Sec.~\ref{SecII}, this occurs because although the large plasmon-coupler detuning
renders static bSWAP-type couplings non-dominant compared to the parametric-activated on-resonance interaction
$|11\rangle\leftrightarrow|22\rangle$, these off-resonance interactions can
still induce temporary populations in noncomputational states. Moreover, the parametric drive introduces
additional contributions to these off-resonance interactions. Note that similar off-resonant
transitions also occur in traditional transmon-based systems. However, in the present
architecture, this issue is more severe because the coupled plasmon-coupler system operates
in a strongly non-dispersive regime.

As shown in Fig.~\ref{fig9}(a), off-resonant transitions generally induce time-dependent population oscillations, with periods and amplitudes determined by the specific transition rate $\Omega_{s}$ and detuning $\Delta_{s}$. This spurious transition thus gives rise to the leakage error (population in a non-computational state) at the end of the gate, given approximately by
\begin{equation}
\begin{aligned}\label{eq11}
L_{1}\propto\frac{\Omega_{s}^2}{\Omega_{s}^2+\Delta_{s}^2} \sin^2\left(\frac{1}{2}\sqrt{\Omega_{s}^2+\Delta_{s}^2}t_{g}\right),
\end{aligned}
\end{equation}
where $t_g$ denotes the gate length. Moreover, during gate operations, off-resonance transitions coexist with the target transition and typically exhibit distinct oscillation periods. Consequently, even when the population oscillation from one interaction is complete, oscillations from the other interactions may remain incomplete, resulting in residual leakage as illustrated in Fig.~\ref{fig9}(a). This leakage can be mitigated by synchronizing the population oscillations of dominant off-resonance transitions (e.g., $|101\rangle\leftrightarrow|121\rangle$) with those of the gate transition (e.g., $|101\rangle\leftrightarrow|202\rangle$)~\cite{Ficheux2021,Economou2015,Barends2019}. In the current architecture with fixed circuit parameters, such synchronization can be achieved by optimizing the flux bias or parametric drive amplitude, both of which affect the on-resonant gate transition rate. More explicitly, for the leading off-resonant transition $|101\rangle\leftrightarrow|121\rangle$, the induced leakage (i.e., the population in $|121\rangle$) can be approximately described by Eq.~(\ref{eq11}), while the population in $|22\rangle$ (arising from the on-resonant gate transition $|11\rangle\leftrightarrow|22\rangle$) at the end of the gate is $P_{22}=\sin^2(\Omega_{d}t_{g})$ (here $\Omega_{d}$ is the gate transition rate). Thus, for $t_{g}=\pi/\Omega_{d}$, the leakage error due to $|101\rangle\leftrightarrow|121\rangle$ is minimized when $\sqrt{\Omega_{s}^2+\Delta_{s}^2}t_{g}=2n\pi$. This condition can be satisfied by optimizing $\Omega_{d}$ (e.g., the parametric drive amplitude) and thus yields multiple optimal gate lengths for suppressing such leakage~\cite{Barends2019}.

The above discussion is mainly qualitative and involves several simplifications, such as assuming square pulses instead of pulses with finite ramp times and neglecting the dependence of the off-resonant transitions on the modulation amplitude. In practical implementations, the synchronization strategy can be realized by scanning the drive parameters (i.e., varying the gate length) to find the optimal settings that minimize leakage, see, e.g., Fig.~\ref{fig7}(b). As demonstrated in Fig.~\ref{fig9}(c) (where the gate length is varied from 45 ns to 65 ns), the synchronization of oscillations significantly reduce both leakage and intrinsic gate error, consistent with the results in Fig.~\ref{fig7}. Moreover, as shown in Fig.~\ref{fig7}, with a shorter gate length (e.g., 40 ns), the leakage error, and hence the gate error, can also be suppressed, as expected. 
 
As the synchronization strategy involves only optimizing the drive amplitude (which is inherently integrated into CZ gate calibration), we expect this scheme to also be applicable for suppressing leakage in large-scale multi-qubit systems. However, we also note that while synchronizing population oscillations between two transitions is generally feasible, extending this synchronization to multiple transitions presents a significant challenge. For example, as shown in Fig.~\ref{fig9}(a), three off-resonance transitions coexist: $|101\rangle\leftrightarrow|211\rangle$, $|101\rangle\leftrightarrow|112\rangle$,
and $|101\rangle\leftrightarrow|121\rangle$. In fast gates employing strong parametric drives, the strengths of
these transitions can become comparable, causing the synchronization approach to fail in suppressing leakage
errors to a negligible level. Since the coupler may also be excited under these conditions, a leakage reduction
operation must address both fluxonium plasmon modes and coupler excitations~\cite{Yang2024}, particularly in the
context of quantum error correction. This added complexity necessitates further optimization of circuit
parameters, for example by increasing the anharmonicity of the transmon coupler, to effectively suppress
these leakage channels.

\subsection{Sensitivity analysis for gate parameter fluctuations}\label{SecIVC}

\begin{figure}[htbp]
\begin{center}
\includegraphics[keepaspectratio=true,width=\columnwidth]{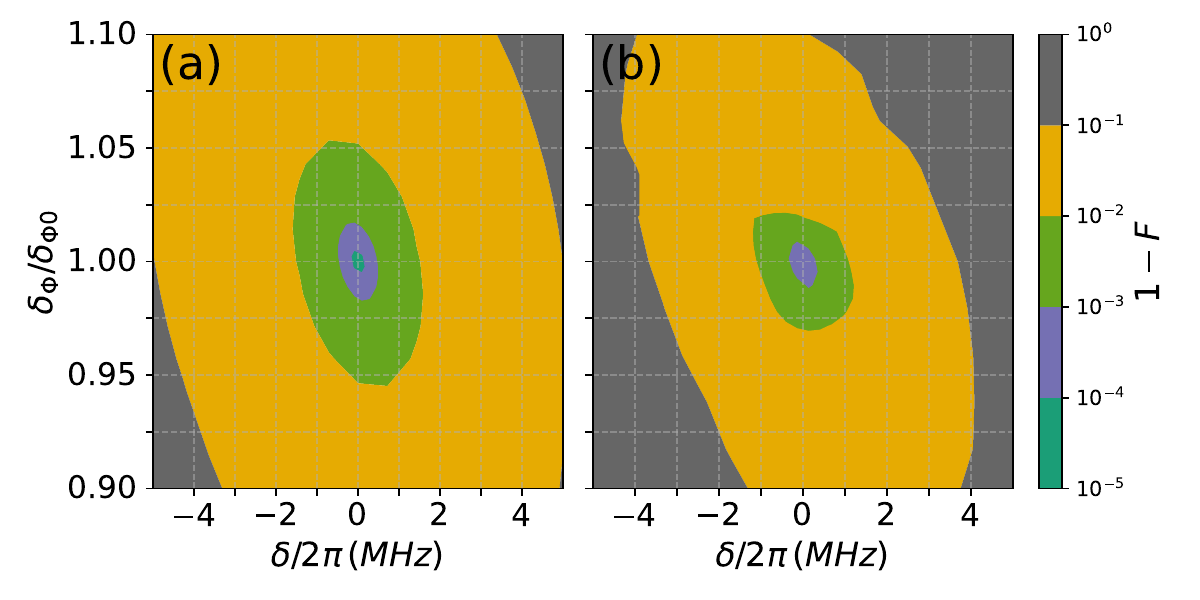}
\end{center}
\caption{Intrinsic error for the 60 ns CZ gate (a) and the 80 ns CZ gate (b) in the two operational
configurations versus drive-frequency detuning $\delta$ and normalized drive amplitude
$\delta_{\Phi}/\delta_{\Phi 0}$ relative to their optimal values. We note that, at the optimal control
points, the intrinsic errors are $6.9\times 10^{-6}$ and $1.9\times 10^{-4}$, respectively, as shown
in Figs.~\ref{fig7}(a) and~\ref{fig7}(c).}
\label{fig10}
\end{figure}

\begin{figure}[htbp]
\begin{center}
\includegraphics[keepaspectratio=true,width=\columnwidth]{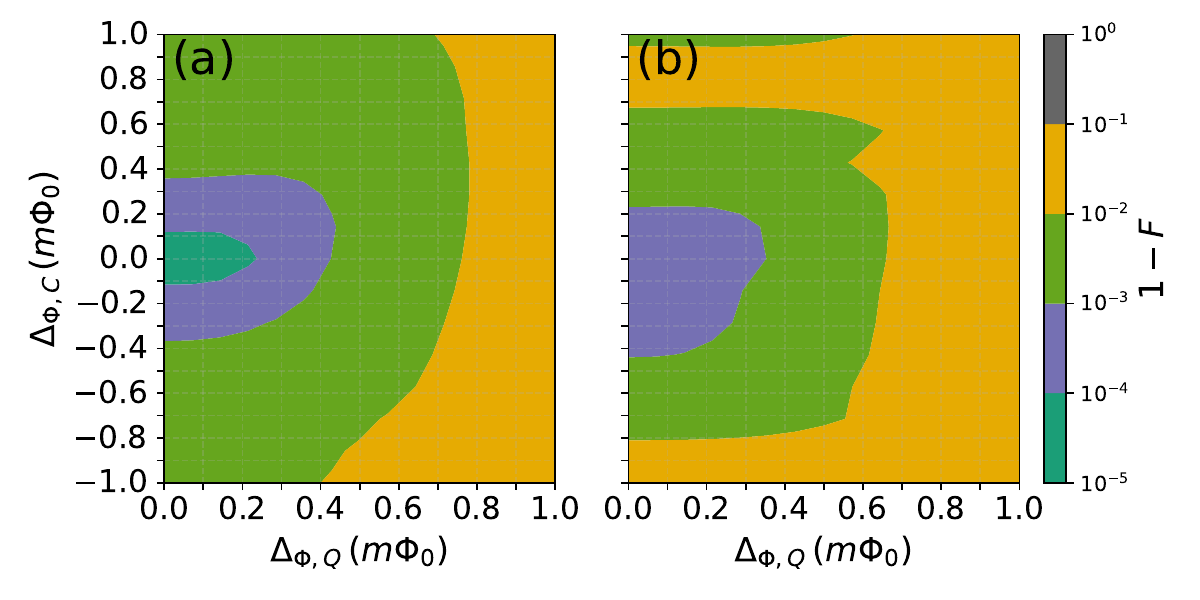}
\end{center}
\caption{Intrinsic error for the 60 ns CZ gate (a) and the 80 ns CZ gate (b) in the two operational
configurations versus qubit-bias offset $\Delta_{\Phi,Q}$ (applied to $Q_{0}$) and
coupler-bias offset $\Delta_{\Phi,C}$ relative to their optimal values. We note that at the optimal control points, the
intrinsic errors are $6.9\times 10^{-6}$ and $1.9\times 10^{-4}$, respectively, as
shown in Figs.~\ref{fig7}(a) and~\ref{fig7}(c).}
\label{fig11}
\end{figure}

Fluctuations or drifts in system parameters (including those of both qubits and control electronics) can degrade gate
performance (see, e.g., Ref.~\cite{Ficheux2021,Ganzhorn2020}). To evaluate their effect on intrinsic gate error, we analyze the sensitivity of the gate implementations to instabilities in various control 
parameters. 

As an illustration, Figures~\ref{fig10} and~\ref{fig11} present results for 60-ns and 80-ns CZ gates in the two operational configurations (using a 5-ns drive-pulse ramp time). Note that, at the optimal control points, the intrinsic gate errors at the optimal control points are $6.9\times 10^{-6}$ and $1.9\times 10^{-4}$, respectively, as shown in Figs.~\ref{fig7}(a) and~\ref{fig7}(c). Based on these optimal parameters, we benchmark CZ gates with intentionally offset parameters and study the resulting gate errors. Figure~\ref{fig10} shows the influence of drive-parameter (frequency and amplitude) fluctuations on the gates, while Fig.~\ref{fig11} displays the effect of bias-parameter (flux biases on fluxonium $Q_{0}$ and the transmon coupler) fluctuations. As expected, fluctuations in system parameters do affect gate performance. At the same time, the results in these figures demonstrate that coherent gate errors below 0.001 are achievable with current experimental technology~\cite{Ficheux2021}.

\subsection{Spectator-induced gate error}\label{SecIVD}

\begin{figure}[htbp]
\begin{center}
\includegraphics[keepaspectratio=true,width=\columnwidth]{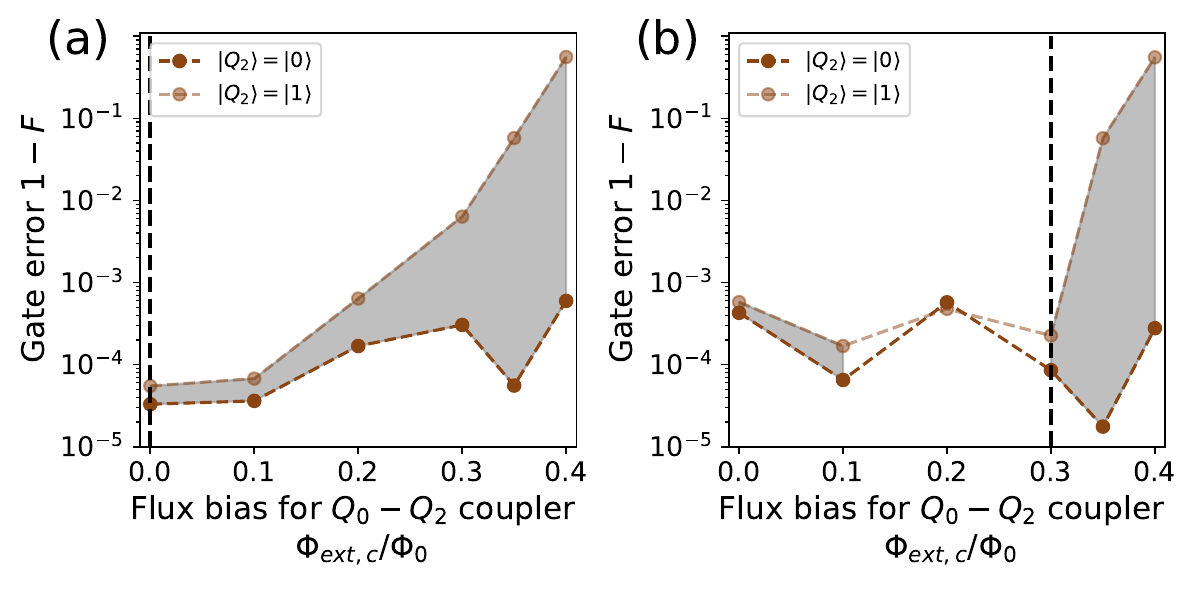}
\end{center}
\caption{Spectator-induced gate errors for the CZ gate applied to
$Q_{0}$ and $Q_{1}$ as a function of the residual coupling between $Q_{0}$ and spectator
fluxonium $Q_{2}$ (controlled via the bias on the $Q_{0}-Q_{2}$ coupler; see
Fig.~\ref{fig1}(a)). Vertical dashed lines mark the bias points that suppress the residual
coupling (see also in Fig.~\ref{fig2}(a)) for suppressing the residuals. (a) 55-ns CZ gates in
the dynamic flux-bias configuration with the spectator $Q_{2}$ prepared in different qubit
states. (b) 75-ns CZ gates in the static flux-bias configuration with the
spectator $Q_{2}$ prepared in different qubit states.}
\label{fig12}
\end{figure}

To assess the extensibility of the proposed gate implementations, we examine the CZ gate
between $Q_{0}$ and $Q_{1}$ in the presence of a spectator qubit $Q_{2}$. As shown in
Fig.~\ref{fig1}(a), the spectator qubit $Q_{2}$ is coupled to $Q_{0}$, with the coupling
strength controlled by the flux bias applied to the $Q_{0}$-$Q_{2}$ coupler. The
three-fluxonium system can be described by the following Hamiltonian
\begin{equation}
\begin{aligned}\label{eq12}
\hat H_{\rm F}=& \sum_{k=0,1,2}[4 E_{C,k} \hat n^2_k + \frac{E_{L,k}}{2}(\hat\varphi_k - \varphi_{\text{ext},k})^2 - E_{J,k}\cos\hat\varphi_k]\\
&+[J_{0}\hat n_0 \hat n_{c0}+J_{0}\hat n_1 \hat n_{c0}+J_{01}\hat n_0 \hat n_{1}]
\\&+[J_{1}\hat n_0 \hat n_{c1}+J_{1}\hat n_3 \hat n_{c1}+J_{03}\hat n_0 \hat n_{3}]
\\&+\sum_{j=0,1}[4 E_{C,cj} \hat n^2_{cj} - E_{J,cj}\cos(\frac{\varphi_{\text{ext},cj}}{2})\cos\hat\varphi_{cj}],
\end{aligned}
\end{equation}
where the subscripts $cj$ denote the two couplers, and $J_{0}$ and $J_{1}$ represent the coupling strengths between the
respective fluxonium qubits and the couplers. The full system parameters utilized here is listed in
Table~\ref{tab:circuit_parameters}. Following typical experimental protocols~\cite{Krinner2020,Cai2021}, the CZ gate
between $Q_{0}$ and $Q_{1}$ is first tuned with spectator fluxonium $Q_{2}$ in $|0\rangle$ and then characterized for
both $|0\rangle$ and $|1\rangle$ states.

For both operational configurations, Figure~\ref{fig12} shows the intrinsic gate error in the presence of residual
coupling between $Q_{0}$ and the spectator fluxonium $Q_{2}$. As expected, stronger $Q_{0}-Q_{2}$ coupling generally increases the gate error for both spectator states, with a more pronounced degradation when the spectator is in $|1\rangle$. Significantly, when the qubit-spectator coupling is suppressed, the gate fidelity approaches that of an isolated two-qubit system, and the gate performance exhibits little dependence on
the spectator state. Conversely, strong coupling can lead to frequency collisions and conditional frequency
shifts (see Fig.~\ref{fig2}(a) and Ref.~\cite{Zhao2025}), which substantially degrade the gate and introduce
spectator-dependent correlated errors. These results, consistent with the analysis in Ref.~\cite{Zhao2025}, underscore
the key role of tunable plasmon interactions in mitigating spectator-induced errors.

\begin{table*}[htbp]
\centering
\caption{Advantages and limitations of existing fluxonium qubit architectures.}
\begin{tabular}{lllll}
\hline\hline 
\textbf{Interaction}          & \textbf{Coupling circuit} & \textbf{Gate state} & \textbf{Advantages} & \textbf{Limitations} \\ \hline        
\multirow{3}{*}{Always-on} & Capacitive & Computational state~\cite{Nguyen2019,Dogan2023,Nesterov2021} & Long coherence; Low leakage & Spectator error\\ 
                           
                           & Capacitive & Non-computational state~\cite{Nesterov2018,Ficheux2021,Simakov2023,Ding2023} & Fast speed 
                           & Short coherence; Spectator error \\ \cline{2-5}
                           
                           & Inductive & Computational state~\cite{Lin2025} & Long coherence; Low leakage & Flux crosstalk; Spectator error\\                        
                            
\\ \hline                           
                           
\multirow{3}{*}{Tunable}   & Capacitive & Computational state~\cite{Moskalenko2021,Moskalenko2022} & 
                            Long coherence; Low leakage & Slow gate speed; ZZ crosstalk \\ 

                           & Capacitive & Non-computational state~\cite{Zhao2025,Zhan2026} (\textbf{This work}) & Fast speed; Low spectator error & Short coherence \\ \cline{2-5}
                           
                           & Inductive & Computational state~\cite{Zhang2024} & Long coherence; Low leakage & Flux crosstalk\\     
                                                
                           \hline\hline                                   
\end{tabular}
\label{tab:gate_architectures}
\end{table*}

\section{Discussion}\label{SecV}

While significant progress has been made in improving the performance of fluxonium-based 
systems (with qubit lifetimes beyond milliseconds~\cite{Somoroff2023} and two-qubit 
gate errors below 0.001~\cite{Ding2023,Zhang2024,Lin2025}), these demonstrations have 
been limited to small-scale processors, typically comprising only two qubits. Thus, attention 
must now turn to scaling fluxonium qubits to large-scale systems if they are to compete 
with transmon-based platforms, which now achieve a median two-qubit gate error below 0.005 
and a lowest gate error below 0.001 in systems at the 100-qubit 
scale (see, e.g., Ref.~\cite{Acharya2025}). As noted earlier, and as highlighted by the development 
of transmon-based systems, addressing challenges such as spectator error (frequency crowding~\cite{Brink2018,Chen2014,Kelly2015} and spurious couplings~\cite{Chen2014,Kelly2015,Mundada2019,Muller2019}), control crosstalk~\cite{Barends2014}, and fabrication variability~\cite{Kreikebaum2020,Hertzberg2021,Pappas2024} is critical for scaling any qubit architecture to large-scale systems. In this context, here we examine the limitations and opportunities of the proposed gate architecture compared with existing fluxonium-based approaches.

Table~\ref{tab:gate_architectures} briefly summarizes the advantages and limitations of existing fluxonium qubit architectures. More explicitly, fluxonium qubits connected via always-on couplings (whether capacitive~\cite{Nguyen2019,Dogan2023,Nesterov2018,Ficheux2021,Simakov2023,Ding2023} or inductive~\cite{Lin2025}) suffer from spectator-induced errors (e.g., ZZ crosstalk~\cite{Nguyen2019,Dogan2023,Ficheux2021}, conditional frequency shifts~\cite{Ding2023,Zhao2025}, and frequency collisions~\cite{Nguyen2019,Nesterov2018}), making such architectures impractical for large-scale systems. As with transmon qubits, this challenge motivates the development of tunable couplers for fluxonium qubits. While tunable control of interactions within the computational subspace has been demonstrated for capacitively coupled systems~\cite{Moskalenko2021,Moskalenko2022}, the fluxonium's weak electric transition dipole moment in that subspace ($< 0.1$, see Table~\ref{tab:transition_parameters}) results in inefficient 
coupling, compromising both gate speed and ZZ suppression. Although this issue can be mitigated by using tunable inductive couplers~\cite{Zhang2024}, flux crosstalk and the small size of the qubit loop make multiple connections impractical for large-scale systems. Consequently, the community has shifted focus to tunable control of plasmon interactions~\cite{Zhao2025}, as we also study here, enabling gate implementations based on microwave-activated transitions to non-computational states. Very recently, such an architecture has been demonstrated experimentally, and the lowest gate error has been found to be below 0.001~\cite{Zhan2026}.

While previous studies have focused on the microwave-activated scheme~\cite{Zhao2025,Zhan2026}, here we turn to the parametric-activated scheme. Since the flxuonium's non-computational states have short coherence times (currently on the order of $\sim10\,\mu s$~\cite{Ficheux2021,Ding2023}) compared with its computational states (which can reach 1 ms~\cite{Somoroff2023}), this poses the dominant error for two-qubit gates in such architectures. As mentioned before, since our approach involves doubly excited plasmon states, the resulting incoherence error is expected to be more severe than in microwave-activated schemes, which typically use only singly excited plasmon states. Fortunately, the  CZ gate length can be commonly suppressed below 100 ns or even below 50 ns, thus gate error approaching 0.001 could still be achieved (see Figs.~\ref{fig7} and~\ref{fig8}). Moreover, as mentioned before, since the decoherence mechanisms of these transitions are similar to those of transmon qubits~\cite{Ficheux2021,Nguyen2019}, one can anticipate further improvements in coherence times, which would push gate errors toward the $10^{-4}$ level. 

Additionally, given the state-of-the-art technology (see Ref.~\cite{Wang2025,Zhan2026}), fluxonium quantum processors are indeed still challenged by fabrication uncertainties in Josephson junctions, which cause deviations 
from the target Josephson energy and inductive energy. Consequently, similar to the microwave-activated 
CZ schemes, our gate architecture also relies on engineering the interactions between plasmon 
transitions and can therefore also be affected by fabrication-induced offsets in the target plasmon transition frequencies. However, the parametric-activated scheme studied here primarily requires that the modulation frequency 
equals the sum of the two fluxonium plasmon transition frequencies, with no constraints on the 
detuning between two coupled fluxonium qubits. In this sense, compared to microwave-activated CZ 
schemes, which require the plasmon–plasmon detuning to lie within a finite range for successful 
gates~\cite{Nesterov2018,Chow2013,Zhao2026b}, the studied gate architecture imposes less stringent requirements on system parameters (e.g., transition frequencies), as also noted in Refs.~\cite{Roth2017,McKay2016} for transmon-based architectures. This could offer better resilience to fabrication variations and parameter misalignments. 
Moreover, the introduction of the tunable coupler, which supports tunable control over plasmon 
interactions, should also provide an additional control knob for compensating potential parameter 
misalignments, compared to architectures with always-on interactions.

\section{conclusion and outlook}\label{SecVI}

In conclusion, we propose a control strategy for fast entangling gates on scalable fluxonium
architectures through parametric modulation of plasmon interaction. For the
parametric-driven system, we identify and categorize three main types of parametric-activated transitions
and demonstrate that, owing to the strong anharmonicity and weak qubit transition dipoles of fluxonium, bSWAP-type plasmon interactions with the strength above $10\,\rm MHz$ can be achieved in two typical operational configurations
without being significantly affected by nearby spurious transitions. This enables the implementation of
sub-100ns CZ gates with intrinsic errors below $10^{-4}$ in both configurations. Furthermore, this parametric
modulation strategy should also be extended to implement native multi-controlled phase gates~\cite{Zhao2025b}, further highlighting the operational flexibility in this architecture.

Similar to the microwave-based approach~\cite{Nesterov2018,Ficheux2021,Xiong2022,Ding2023,Zhao2025}, the present technique also faces specific challenges. We show that since parametric-activated CZ gates involve temporary occupation of noncomputational states, the dominant gate error in practical devices is expected to arise from relaxation and dephasing of doubly excited plasmon states. Given current coherence
times on the order of $10\ \mu\text{s}$~\cite{Ficheux2021,Ding2023}, achieving gate errors
approaching $10^{-4}$ requires further improvement of plasmon coherence, which may be realized by optimizing
chip fabrication, fluxonium parameters, and device layout in further implementations. Additionally, since leakage
into noncomputational states, particularly those involving coupler excitation, may be unavoidable, leakage
reduction operations for both fluxonium and coupler excitations will be necessary in the context of quantum
error correction~\cite{Miao2023}. Moreover, as the inductive shunt in a fluxonium circuit is typically
implemented using a Josephson junction array~\cite{Manucharyan2009,Nguyen2019,Somoroff2023,Wang2025}, the internal degrees of freedom of this array can give rise to parasitic array modes~\cite{Sorokanich2024}, which may strongly couple to the fluxonium plasmons. Therefore, in the current fluxonium system under parametric drives, device parameters must be carefully designed to prevent the activation of unwanted interactions involving these parasitic modes. Overall, driven by the pursuit of high system-level performance for quantum error correction and considering the complex
spectral characteristics of fluxonium-based architectures, these requirements call for a broader architectural
perspective that balances various potential issues through careful design trade-offs to enable scalable
fluxonium quantum processors.

\begin{acknowledgments}
Peng Zhao would like to thank Zhuang Ma for insightful discussions.
This work is supported by the National Natural Science Foundation of China (Grants No.12204050 and No.92576110) and the Guangdong Provincial Quantum
Science Strategic Initiative (Grant No. GDZX2203001). Peng Xu is supported by the National Natural Science
Foundation of China (Grants No.12105146, No.12175104, and No.92565111) and the Program of State Key Laboratory of Quantum Optics
Technologies and Devices (No.KF202505).
\end{acknowledgments}

\appendix

\section{Effective system Hamiltonian}\label{A}

Following the approach in Ref.~\cite{Zhao2025}, we derive the effective system Hamiltonian presented in the
main text. We consider a system composed of two fluxonium qubits, $Q_{0}$ and $Q_{1}$, coupled via a transmon-based
tunable coupler. The full Hamiltonian of the system is given by:
\begin{equation}
\begin{aligned}\label{eqA1}
\hat{H}=& \sum_{k=0,1}[4 E_{C,k} \hat n^2_k + \frac{E_{L,k}}{2}(\hat\varphi_k - \varphi_{\text{ext},k})^2 - E_{J,k}\cos\hat\varphi_k]\\
&+J_{c0}\hat n_0 \hat n_{c}+J_{c1}\hat n_1 \hat n_{c}+J_{01}\hat n_0 \hat n_{1}
\\&+4 E_{C,c} \hat n^2_{c} - E_{J,c}\cos(\frac{\varphi_{\text{ext},c}}{2})\cos\hat\varphi_{c}.
\end{aligned}
\end{equation}
By approximating the transmon coupler as an anharmonic oscillator~\cite{Koch2007} and introducing
\begin{equation}
\begin{aligned}\label{eqA2}
&\hat \varphi_{c} = \phi_{c,{\rm zpf}}(\hat a_{c}^{\dag}+\hat a_{c}),
\quad \hat n_{c} = i n_{c,{\rm zpf}}(\hat a_{c}^{\dag}-\hat a_{c})
\end{aligned}
\end{equation}
with
\begin{equation}
\begin{aligned}\label{eqA3}
&\varphi_{c,{\rm zpf}} =\frac{1}{\sqrt{2}}\left[\frac{8E_{C,c}}{E_{J,c}(\varphi_{\text{ext},c})}\right]^{\frac{1}{4}},
n_{c,{\rm zpf}}= \frac{1}{\sqrt{2}}\left[\frac{E_{J,c}(\varphi_{\text{ext},c})}{8E_{C,c}}\right]^{\frac{1}{4}},
\end{aligned}
\end{equation}
the coupler Hamiltonian can be approximated by
\begin{equation}
\begin{aligned}\label{eqA4}
\hat{H}_{coupler}=\omega_{c}\hat a_{c}^{\dag}\hat a_{c}+\frac{\eta_{c}}{2}\hat a_{c}^{\dag}\hat a_{c}^{\dag}\hat a_{c}\hat a_{c},
\end{aligned}
\end{equation}
where $a_{c}$ ($a_{c}^{\dag}$) denotes the destroy (creation) operator, $\phi_{c,{\rm zpf}}$ ($n_{c,{\rm zpf}}$) represents
the phase (number) zero-point fluctuation, and $\omega_{c}$ and $\eta_{c}$ are the transition frequency and the anharmonicity
of the coupler, respectively.

When focusing on a specific plasmon mode per fluxonium, e.g., $|j\rangle \rightarrow |l\rangle$ in $Q_0$
and $|r\rangle \rightarrow |t\rangle$ in $Q_1$, we define the corresponding lowering and raising operators as follows:
\begin{equation}
\begin{aligned}\label{eqA5}
&\hat p_{0}=|j\rangle\langle l|,\,\hat p_{0}^{\dag}=|l\rangle\langle j|,
\\&\hat p_{1}=|r\rangle\langle t|,\,\hat p_{1}^{\dag}=|t\rangle\langle r|,
\end{aligned}
\end{equation}
for the plasmon modes of the two fluxoniums, with transition frequencies $\omega_{p,0}$ and $\omega_{p,1}$, respectively.

Accordingly, the full system Hamiltonian takes the following form after redefining $a=-ia$ ($a^{\dag}=ia^{\dag}$)
\begin{equation}
\begin{aligned}\label{eqA6}
\hat H_{p}=&\sum_{k=0,1}\left[\omega_{p,k}\hat p_{k}^{\dag}\hat p_{k}+{\color{blue}g_{p,k}}(\hat p_{k}+\hat p_{k}^{\dag})(\hat a_{c}+\hat a_{c}^{\dag})\right]
\\&+\omega_{c}\hat a_{c}^{\dag}\hat a_{c}+\frac{\alpha_{c}}{2}\hat a_{c}^{\dag}\hat a_{c}^{\dag}\hat a_{c}\hat a_{c}
+g_{p,01}(\hat p_{0}+\hat p_{0}^{\dag})(\hat p_{1}+\hat p_{1}^{\dag}),
\end{aligned}
\end{equation}
where
\begin{equation}
\begin{aligned}\label{eqA7}
&g_{p,0}=J_{c0}\langle j|\hat n_0| l\rangle \langle 1 |\hat n_{c}|0\rangle,
\\&g_{p,1}=J_{c1}\langle r|\hat n_1| t\rangle \langle 1 |\hat n_{c}|0\rangle,
\\&g_{p,01}=J_{01}\langle j |\hat n_0|l\rangle \langle t|\hat n_{1}|r\rangle,
\end{aligned}
\end{equation}
represent the coupling strengths of the plasmon-coupler couplings and the direct plasmon-plasmon
coupling. The magnitudes of the transition matrix elements for the plasmon modes and the coupler used
in this work are summarized in Table~\ref{tab:transition_parameters}.

Considering that the coupled plasmon-transmon system operates in the dispersive regime, i.e., the interaction
strength $g_{p,k}$ significantly smaller than the plasmon-coupler detuning $\Delta_{p,k}=|\omega_{p,k}-\omega_{c}|$, an
effective Hamiltonian can be obtained by eliminating the direct plasmon-coupler interactions~\cite{Bravyi2011,Zueco2009},
leading to (up to the second order in $g_{p,k}/\Delta_{p,k}$)
\begin{equation}
\begin{aligned}\label{eqA8}
\hat H_{p,{\rm eff}}=&\sum_{k=0,1}\left[(\omega_{p,k}+\frac{g_{p,k}^2}{\Delta_{p,k}})\hat p_{k}^{\dag}\hat p_{k}\right]
\\&+\left(\omega_{c}-\sum_{k=0,1}\frac{g_{p,k}^2}{\Delta_{p,k}}\right)\hat a_{c}^{\dag}\hat a_{c}+\frac{\alpha_{c}}{2}\hat a_{c}^{\dag}\hat a_{c}^{\dag}\hat a_{c}\hat a_{c}
\\&+ g_{p}(\hat p_{0}+\hat p_{0}^{\dag})(\hat p_{1}+\hat p_{1}^{\dag}).
\end{aligned}
\end{equation}
Here, the final term represents the coupler-mediated plasmon-plasmon interaction, with strength
\begin{equation}
\begin{aligned}\label{eqA9}
g_{p}=g_{p,01}+\frac{g_{p,0}g_{p,1}}{2}\left[\sum_{k=0,1}(\frac{1}{\Delta_{p,k}}-\frac{1}{S_{p,k}})\right].
\end{aligned}
\end{equation}
and $S_{p,k}=\omega_{p,k}+\omega_{c}$.

\begin{figure*}[tbp]
\begin{center}
\includegraphics[width=16cm,height=12cm]{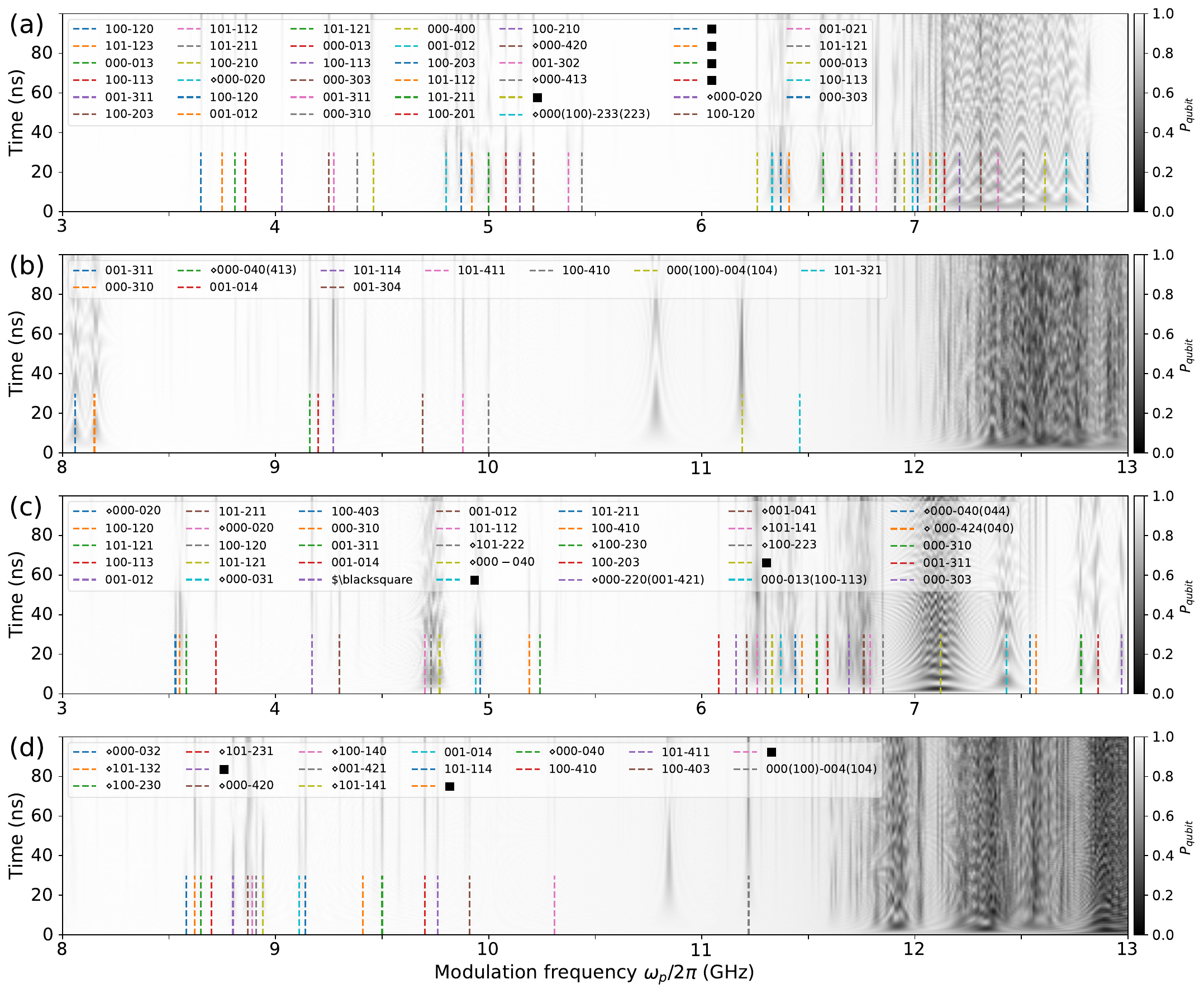}
\end{center}
\caption{Identical to Fig.~\ref{fig3}, but explicitly illustrating the physical origin of the chevron patterns
observed over the frequency ranges 3–8 GHz in (a,c) and 8–13 GHz in (b,d). Panels (a,b) and (c,d) correspond to the dynamic flux-bias configuration and the static bias configuration, respectively. Apart from the transitions highlighted by the diamonds (which involve excitation of the coupler mode) and the black squares, all other transitions can be described by the categories given in Fig.~\ref{fig14}. Note that the weak harmonic approximation (based on the Fock basis) for the transmon coupler may break down when describing highly excited states. Consequently, results for transitions involving such states, specifically those giving rise to the chevron pattern marked by the black square, should be considered unreliable and a more accurate description requires the use of the charge basis.}
\label{fig13}
\end{figure*}

\section{spurious transition}\label{B}

As noted in the main text, the parametric-driven fluxonium system exhibits three primary classes of state
transitions, \textbf{(1) bSWAP-type transitions for plasmon mode pairs},
\textbf{(2) Blue sideband transitions between the fluxonium plasmon modes and the coupler}, and
\textbf{(3) Coupler state excitations due to effective two-photon (squeezing) drives}. Additionally, cross-resonance-like
transitions arise due to strong state hybridization. The following discussion elaborates on the physical
mechanisms underlying these transitions. For clarity, the following analysis is restricted to the dynamic
flux-bias configuration.

\subsection{The analysis of the spurious transition}\label{B1}

\begin{figure}[tbp]
\begin{center}
\includegraphics[keepaspectratio=true,width=\columnwidth]{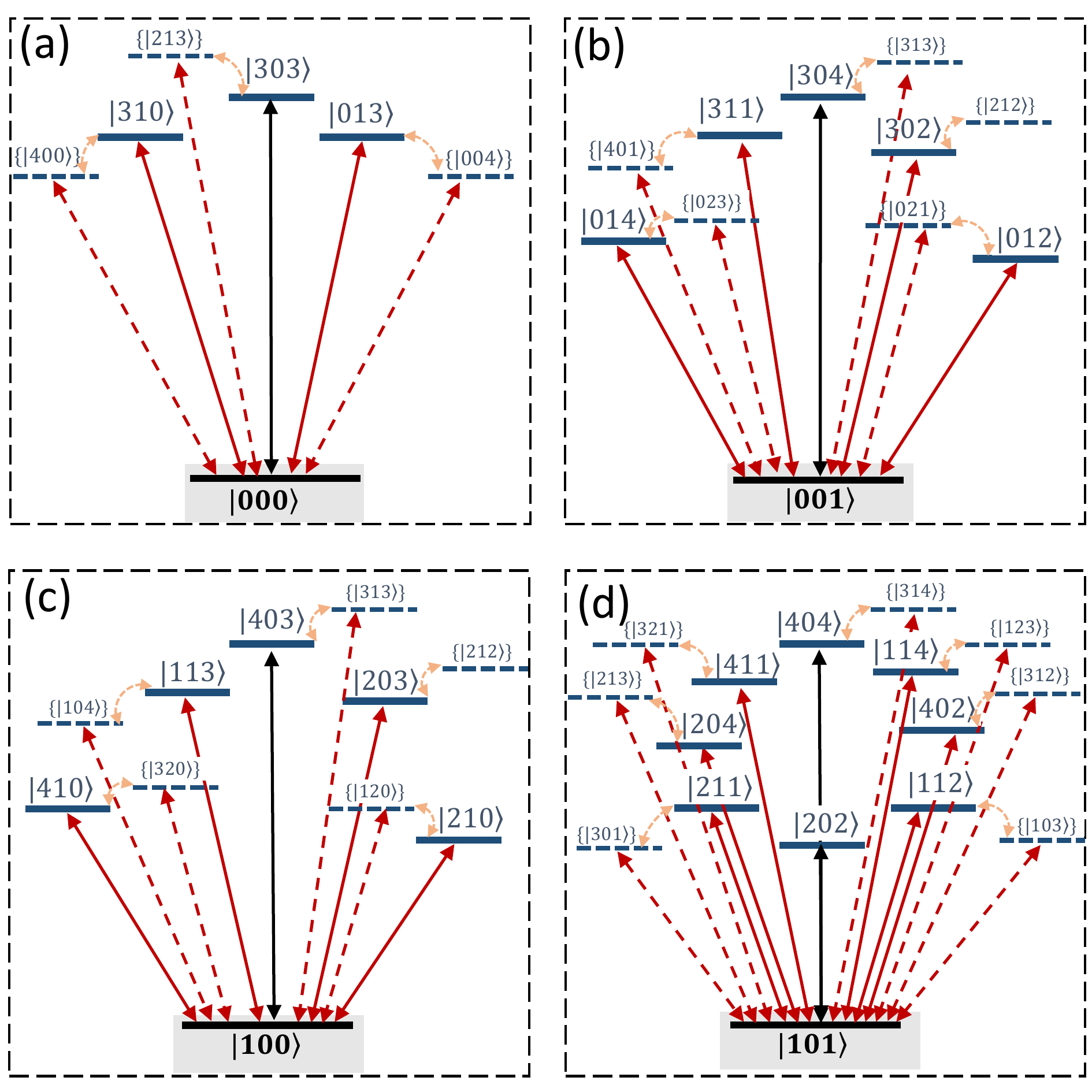}
\end{center}
\caption{A schematic illustrating parametric-activated state transitions between computational and
non-computational states. (a)-(d) correspond to the computational states $|000\rangle$, $|001\rangle$, $|100\rangle$,
and $|101\rangle$, respectively. In addition to the two main transition types, namely bSWAP-type transitions and sideband transitions (see solid black and red lines, respectively), cross-resonance-like
spurious transitions (dashed red lines) are present, resulting from strong state hybridization induced by
plasmon-coupler interactions, as indicated by the orange arrows. Here, $\{|ijk\rangle\}$ denotes a set of states; for example, $\{|103\rangle\}\equiv\{|103\rangle,|121\rangle,...\}$ is the set of states coupled to $|112\rangle$ via plasmon–coupler interactions.}
\label{fig14}
\end{figure}

Similar to Fig.~\ref{fig3}, Figure~\ref{fig13} reveals the specific state transitions responsible
for the chevron patterns observed across the 3-13 GHz frequency range. Corresponding detailed illustrations
of the two main transition types, i.e., bSWAP-type plasmon transitions (solid black lines) and blue sideband transitions (solid red lines), are provided for all four computational states in Fig.~\ref{fig14}. Additionally, the physical mechanism underlying the cross-resonance-like transitions (dashed red lines), facilitated by either bSWAP-type plasmon transitions or sideband transitions, is explicitly elucidated as an example.

\subsection{Effective coupler drive}\label{B2}

Here we provide further details regarding the effective drive experienced by the transmon coupler (the transitions marked by diamonds and black squares in Fig.~\ref{fig13}). As illustrated
in Fig.~\ref{fig13}, the parametric drive can induce transitions within the coupler states or transitions that
involve coupler excitation, such as $|000\rangle\rightarrow|020\rangle$. To elucidate the physical mechanism
underlying these transitions, we examine the case of an isolated transmon coupler subject to parametric drives.

Considering that the SQUID of the frequency-tunable transmon coupler is threaded by a time-dependent magnetic
flux, the Hamiltonian of this flux-driven system takes the following form~\cite{You2019,Riwar2022}:
\begin{equation}
\begin{aligned}\label{eqB1}
&\hat H_{coupler} = 4E_{C,c}\hat{n}_{c}^2+\hat{U}_{c},
\\&\hat{U}_{c}=-E_{J1}\cos{(\hat\varphi_{c}+\alpha\varphi_{{\rm ext},c})}-E_{J2}\cos{(\hat\varphi_{c}+\beta\varphi_{{\rm ext},c})}
\end{aligned}
\end{equation}
where $E_{J1}$ and $E_{J2}$ denote the Josephson energies of the two junctions in the SQUID, and $\alpha$
and $\beta$ are parameters determined by the specific device layout~\cite{Riwar2022}, constrained by the
relation $\alpha-\beta=1$.

Without loss of generality, we assume $E_{J1}=E_{J2}=E_{J,c}/2$, and the external flux takes the
form $\varphi_{{\rm ext},c}=\varphi_{{\rm ext},s}+\varphi_{{\rm ext},d}(t)$, where
$\varphi_{{\rm ext},d}\ll1$. The potential energy of the flux-driven transmon can then be expressed as:
\begin{equation}
\begin{aligned}\label{eqB2}
\hat{U}_{c}/E_{J,c}&=-\cos(\frac{\varphi_{{\rm ext},c}}{2})\cos(\hat\varphi_{c}+d\varphi_{{\rm ext},c}),
\end{aligned}
\end{equation}
where $d=\alpha+\beta$. When equal capacitance is assigned to each Josephson junction in the SQUID (i.e., $d=0$), the
potential Hamiltonian reduces to the form used in the present work, as given in Eq.~(\ref{eqA1}). In this case, expanding the potential in Eq.~(\ref{eqB2}) to second order in $\varphi_{{\rm ext},d}(t)$ ($\hat\varphi_{c}$) yields the following approximate expression:
\begin{equation}
\begin{aligned}\label{eqB3}
\hat{U}_{c}/E_{J,c}\approx & \cos(\frac{\varphi_{{\rm ext},s}}{2})\frac{\hat\varphi_{c}^{2}}{2}-\sin(\frac{\varphi_{{\rm ext},s}}{2})\varphi_{{\rm ext},d}(t)\frac{\hat\varphi_{c}^{2}}{4}
\\&-\cos(\frac{\varphi_{{\rm ext},s}}{2})\varphi_{{\rm ext},d}(t)^{2}\frac{\hat\varphi_{c}^{2}}{8}.
\end{aligned}
\end{equation}

By employing the anharmonic oscillator representation introduced in Appendix~\ref{A}, it can be shown that
the second and third terms in Eq.~(\ref{eqB3}) correspond to an effective two-photon (squeezing)
drive, i.e., $\sim (\hat{a}_{c}\hat{a}_{c}+\hat{a}_{c}^{\dag}\hat{a}_{c}^{\dag})$, on the transmon coupler. This drive operates through both single- and two-photon processes. These terms can fasciate the excitation of the transmon coupler when the parametric drive frequency is resonant with state transitions involving the transmon, such as $|000\rangle\rightarrow|020\rangle$ and the transitions indicated by the diamonds and the black squares, which involves multiple excitations of the coupler (see Fig.~\ref{fig13}). Note that the weak harmonic approximation (based on the Fock basis) for the transmon coupler may become invalid for highly excited states. Therefore, results for transitions involving such states, particularly those producing the chevron pattern marked by the black square, are unreliable and require a more accurate description using the charge basis.

We note that the above analysis assume a special case, where $d=0$. i.e., in the SQUID equal capacitance
is assigned to each Josephson junction. However, for actual devices, the unambiguous capacitance assignment
should take into consideration of the detailed device geometric~\cite{Riwar2022}, which might be beyond the scope of the present work. Here we thus turn to give only the qualitative analysis for the most general case $d\neq 0$.
For clarity, assuming $d\varphi_{{\rm ext},c}\ll1$ in Eq.~(\ref{eqB2}), expanding
the potential in Eq.~(\ref{eqB2}) to second order in $d\varphi_{{\rm ext},c}$ ($\hat\varphi_{c}$) yields the
following approximate expression:
\begin{equation}
\begin{aligned}\label{eqB4}
\hat{U}_{c}/E_{J,c}\approx &\cos(\frac{\varphi_{{\rm ext},c}}{2})\frac{\hat\varphi_{c}^{2}}{2}+\cos(\frac{\varphi_{{\rm ext},c}}{2})d\varphi_{{\rm ext},c}\hat\varphi_{c}
\\&-\cos(\frac{\varphi_{{\rm ext},c}}{2})\frac{d^2}{6}\varphi_{{\rm ext},c}^{2}\hat\varphi_{c}^{2}.
\end{aligned}
\end{equation}

Within the anharmonic oscillator representation framework, it can be shown that, unlike the case with $d=0$, the general scenario $d\neq 0$ introduces not only the two-photon (squeezing) drive $\sim (\hat{a}_{c}\hat{a}_{c}+\hat{a}_{c}^{\dag}\hat{a}_{c}^{\dag})$ (third term) but
also a single-photon coupler drive $\sim (\hat{a}_c + \hat{a}_c^\dag)$ (second term). We therefore note that under strong parametric drive, these additional coupler drive terms may induce ionization of the transmon coupler~\cite{Xia2025,Dumas2025}, which
could serve as another limiting factor for realizing fast, high-fidelity parametric-activated gates.

\section{The implementation of the Parametric-activated CZ Gate}\label{C}

\begin{figure}[htbp]
\begin{center}
\includegraphics[keepaspectratio=true,width=\columnwidth]{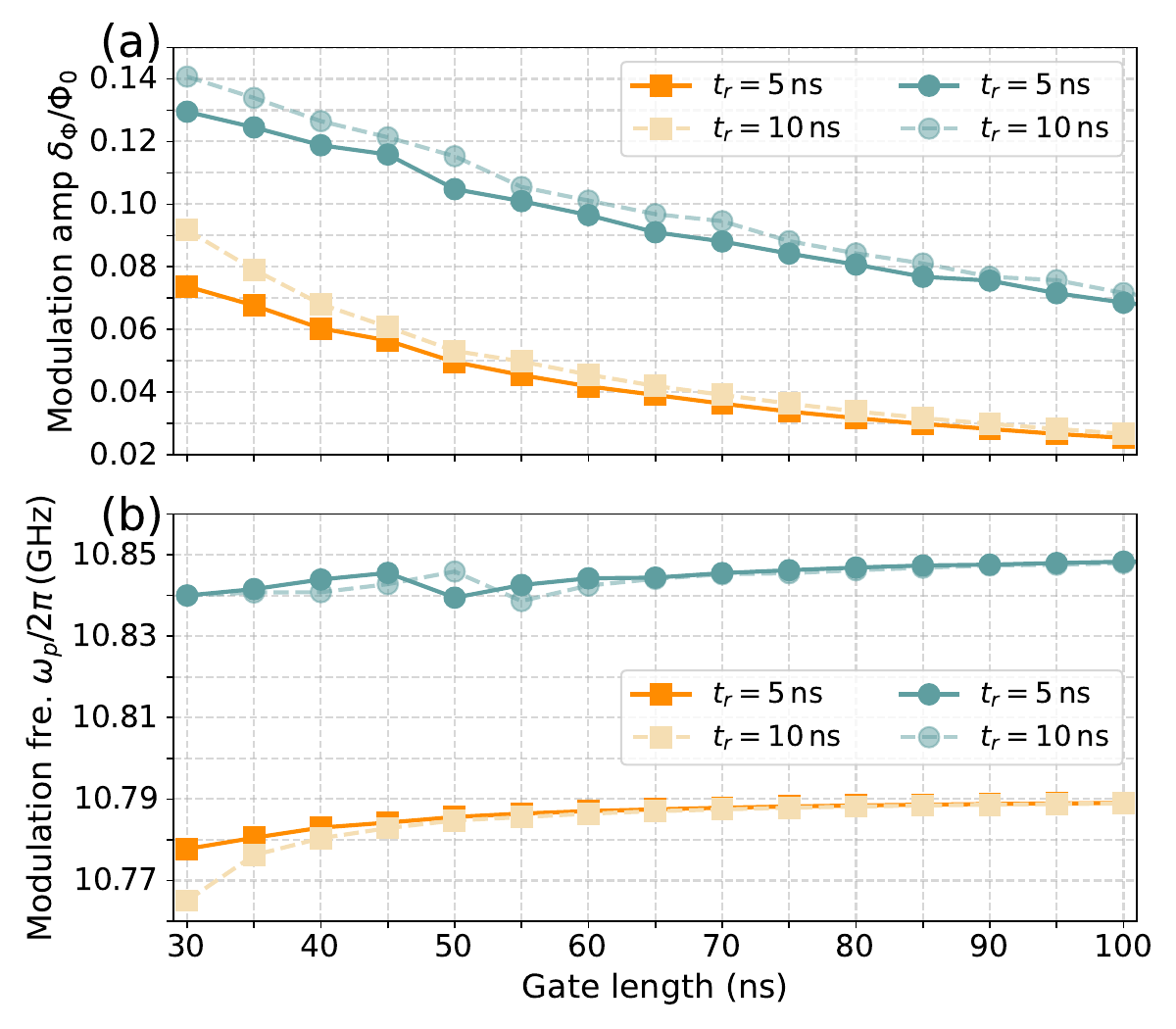}
\end{center}
\caption{The optimized gate parameters (i.e., modulation amplitudes and frequencies) used
for the results shown in Fig.~\ref{fig7}.}
\label{fig15}
\end{figure}

As mentioned in the main text, the CZ gate is realized by applying a parametric drive to the coupler, with the
drive waveform given by:
\begin{equation}
\begin{aligned}\label{eqC1}
\Phi(t)=\Phi_{s}+\Phi_{d}\cos(\omega_{p}t+\phi_{0}),
\end{aligned}
\end{equation}
where $\Phi_{s}$ represents the static bias, $\Phi_{d}$, $\omega_{p}$, and $\phi_{0}$ denotes the amplitude, frequency, and the phase of the parametric drive, respectively. For the parametric drive,  we employ a flat-top cosine pulse defined by
\begin{align}
\Phi_{d}\equiv
\begin{cases}
\delta_{\Phi}\frac{1-\cos{(\pi \frac{t}{t_r}})}{2}  \;, &0<t<t_r\\
\delta_{\Phi}\;,  &t_r<t<t_g-t_r\\
\delta_{\Phi}\frac{1-\cos{(\pi \frac{t_g-t}{t_r}})}{2} \;, &t_g-t_r<t<t_g
\end{cases}
\label{eqC4}
\end{align}
with the ramp time $t_{r}$. A similar pulse shape is also used for the dynamic flux bias in
the dynamic flux-bias configuration.

Following the procedure in Refs.~\cite{Zhao2025b,Jiang2025}, the optimized gate parameters (i.e., the drive amplitude $\delta_{\Phi}$ and the modulation frequency $\omega_{p}$, see Fig.~\ref{fig15}) are determined by minimizing both leakage~\cite{Wood2018} and conditional phase errors within the computational subspace spanned by $\{|00\rangle,\,|01\rangle,\,|10\rangle,\,|11\rangle\}$. Note that similar to the microwave-activated CZ schemes demonstrated in Refs.~\cite{Ding2023,Zhan2026}, the gate tune-up procedure in practical implementations should also focus on minimizing both the conditional phase error and the leakage error 
by optimizing the parametric-drive frequency and amplitude. More explicitly, both the microwave-activated scheme and the parametric-activated scheme studied here rely on activating a complete Rabi oscillation between a computational state (e.g., $|11\rangle$) and a non-computational state (i.e., $|21\rangle$ for the microwave-activated scheme 
and $|22\rangle$ for our case) by optimizing the drive frequency and amplitude. Moreover, the studied coupling architecture, in which fluxonium qubits are coupled via a transmon-based tunable coupler that supports tunable control over plasmon interactions (theoretically studied in Ref.~\cite{Zhao2025} for the microwave-activated CZ scheme), has very recently been realized experimentally and used to demonstrate fast, high-fidelity microwave-activated CZ gates 
in Ref.~\cite{Zhan2026}. Thus, in principle, the gate calibration procedure reported in Ref.~\cite{Zhan2026} should also be applicable to the parametric-activated scheme studied here. We thus expect that no fundamental limitations are likely to hinder gate calibration in practical implementations.

The intrinsic gate performance (excluding decoherence effects) is subsequently evaluated using the state-average gate fidelity metric~\cite{Pedersen2007}, defined as (up to single-qubit phases, which are typically corrected using virtual-Z gates with near-perfect fidelity and zero duration~\cite{McKay2017}):
\begin{equation}
\begin{aligned}\label{eqC5}
F=\frac{{\rm Tr}(\hat{U}^{\dagger}\hat{U})+|{\rm Tr}(\hat{U}_{\rm cz}^{\dag}\hat{U})|^{2}}{N(N+1)},
\end{aligned}
\end{equation}
where $\hat{U}$ represents the truncated actual evolution operator within the computational subspace, $\hat{U}_{\rm cz}$ corresponds to the ideal CZ gate, and $N=4$ is the dimension of the computational subspace for the two-qubit case. The first term quantifies the leakage out of the computational subspace, whereas the second term is the squared modulus of the inner product between the actual evolution operator and the ideal one. Here the actual evolution operator $U$ is obtained through numerical simulation of the gate dynamics, where each fluxonium is truncated to its five lowest energy levels and the transmon coupler is described in a Fock basis~\cite{Koch2007} with a dimension of 21. 

Although the intrinsic CZ gate error is already very low (e.g., on the order of $10^{-4}$), practical
implementations may still be constrained by various error sources, such as the CZ phase error, the drive amplitude error arising from imperfect parameter calibration, and the incoherence error due to relaxation and dephasing of the fluxonium and the coupler. Given current experimental technology~\cite{Ficheux2021,Ding2023}, the dominant gate error is likely due to incoherent processes (for errors arising from imperfect calibration, we refer the reader to the analyses in Ref.~\cite{Ding2023}). In the following, we therefore focus on the incoherence error from relaxation and dephasing of the fluxonium and coupler, treating it both numerically and analytically.

\subsection{Numerical analysis}\label{C1}

Here, the influence of qubit and coupler decoherence on CZ gate performance is investigated using the Lindblad master equation. Without loss of generality, when a decoherence channel $\hat{O}_{j}$ is taken into account, the system dynamics are governed by the master equation, which takes the form
\begin{eqnarray}
\begin{aligned}\label{eqC6}
\frac{d\rho}{dt}=-i[\hat{H},\rho]+\gamma_{j}\left[\hat{O}_{j}\rho \hat{O}_{j}^{\dag}-\frac{1}{2}(\hat{O}_{j}^{\dagger }\hat{O}_{j}\rho+\rho \hat{O}_{j}^{\dagger }\hat{O}_{j})\right],
\end{aligned}
\end{eqnarray}
where the Hamiltonian $H$ is given in Eq.~(\ref{eq1}), $\rho$ is the density matrix of the system, $\gamma_{j}$ denotes the decoherence rate associated with the channel $\hat{O}_{j}$. Here, the first term describes the coherent dynamics governed by the Hamiltonian $H$, whereas the second term describes the incoherent part of the dynamics induced by the decoherence channel $\hat{O}_{j}$. In our numerical simulations, we include all relevant
decoherence channels as dictated by the above master equation, encompassing relaxation and dephasing of
both the fluxonium and the coupler.

Here, because the fidelity metric defined in Eq.~(\ref{eqC5}) is not applicable under decoherence, we adopt the average
gate fidelity defined as~\cite{Wood2018,Zhao2022c}
\begin{eqnarray}
\begin{aligned}\label{eqC7}
F=\frac{N\,F_{p}+1-L_{1}}{N+1},
\end{aligned}
\end{eqnarray}
where $L_{1}$ denotes the leakage of the gate operation~\cite{Wood2018} and $F_{p}$ is the process fidelity
of the implemented CZ gate. Both quantities are obtained by numerically solving the above Lindblad master
equation~\cite{Wood2018,Zhao2022c}. Here, we assume the computational states ($|0\rangle-|1\rangle$)
of the fluxonium have relaxation and dephasing times of $T_{1}=1\,\rm ms$ and $T_{\phi}=0.2\,\rm ms$, consistent
with recent experiments (Ref.~\cite{Ding2023}). For the fluxonium plasmon transitions (e.g., $|1\rangle-|2\rangle$,
$|0\rangle-|3\rangle$, $|1\rangle-|4\rangle$) and the coupler transitions, we set $T_{1}=T_{2}$. Besides, we further
assume the fluxonium plasmon modes have a relaxation time half that of the transmon coupler.

Note that unlike the calculation of intrinsic gate errors, the transmon coupler
is here modeled as a three-level anharmonic oscillator (see Ref.~\cite{Zhao2025b}) to lower the computational
overhead of solving the Lindblad equation, while still capturing the essential physics of
the coupled fluxonium system (the same modeling approach is also used in the analysis of spectator-induced
gate errors). This difference in modeling leads to distinct intrinsic gate errors
for CZ gates of the same duration, i.e., the 55-ns and 75-ns CZ gates in the two operational
configurations, as illustrated in Figs.~\ref{fig7} and~\ref{fig8}.

\subsection{Analytical analysis}\label{C2}

Here, we present an analytical treatment of the incoherence errors in the current gate scheme. For practical high-coherence fluxonium systems~\cite{Ficheux2021,Ding2023}, the dominant incoherent error is expected to arise from relaxation and dephasing of the non-computational gate states~\cite{Ficheux2021,Ding2023,Abad2023}, which is also numerically verified in Fig.~\ref{fig8}. We therefore focus on the relaxation and dephasing of the fluxonium's $|2\rangle$ state. Note that while we do not directly study the incoherence error from coupler relaxation and dephasing, their effect on the gate scheme is implicitly taken into account through the analysis of the decoherence of the fluxonium's $|2\rangle$ state, to which coupler relaxation and dephasing contribute (i.e., the coupler-induced Purcell decay and the coupler-induced dephasing, we refer the reader to the analyses in Ref.~\cite{Zhao2025}).

Since our scheme relies on the parametrically activated gate transition $|11\rangle\leftrightarrow|21\rangle$, 
the evolution operator for the CZ gate can be approximated by
\begin{equation}
\begin{aligned}\label{eqF5}
\hat{U}_{\rm gate}(t)=&|00\rangle\langle 00|+|01\rangle\langle 01|+|10\rangle\langle 10|\\
&+|21\rangle\langle 21|+|12\rangle\langle 12|\\
&+\cos(\Omega t)(|11\rangle\langle 11|+|22\rangle\langle 22|)\\
&-i\sin(\Omega t)(|11\rangle\langle 22|+|22\rangle\langle 11|),
\end{aligned}
\end{equation}
where $\Omega$ denotes the activated transition rate of $|11\rangle\leftrightarrow|21\rangle$ and $\Omega t_{g}=\pi$. Within the framework of the Lindblad master equation, the relaxation and dephasing
of the two fluxonium's $|2\rangle$ states can be described by following four collapse operators
\begin{equation}
\begin{aligned}\label{eqF6}
&\hat{L}_{1r}=\sqrt{\frac{1}{T_{1}^{21}}}|1\rangle\langle 2|\otimes \hat{I},\quad \hat{L}_{1\phi}=\sqrt{\frac{2}{T_{\phi}^{21}}}|2\rangle\langle 2|\otimes\hat{I},
\\&\hat{L}_{2r}=\sqrt{\frac{1}{T_{1}^{21}}}\hat{I}\otimes |1\rangle\langle 2|,\quad \hat{L}_{2\phi}=\sqrt{\frac{2}{T_{\phi}^{21}}}\hat{I}\otimes |2\rangle\langle 2|,
\end{aligned}
\end{equation}
where $\hat{I}$ is the identity operator, and $T_{1}^{21}$ and $T_{\phi}^{21}$ denote the relaxation time and the (white-noise) dephasing time of the fluxonium's $|2\rangle$ state, respectively. For illustration, we here assume identical relaxation and dephasing times for both fluxonium qubits.

Following Ref.~\cite{Abad2023}, the incoherent error due to the collapse operator $\hat{L}$ can
be approximated by
\begin{equation}
\begin{aligned}\label{eqF7}
\epsilon_{\hat{L}} =& \int_{0}^{t_g}dt\left(\frac{{\rm Tr}[\hat{L}^\dag(t)\hat{L}(t)]}{5}-\frac{{\rm Tr}[ \hat{L}^\dag(t)] {\rm Tr}[ \hat{L}(t)]}{20}
\right)
\end{aligned}
\end{equation}
where $\hat{L}(t)=\hat{U}_{\rm gate}^{\dag}(t)\hat{L}\hat{U}_{\rm gate}(t)$ and the trace runs only over the computational subspace. Consequently, the incoherent gate errors from the relaxation and dephasing of the fluxonium's $|2\rangle$ state can be expressed as
\begin{equation}
\begin{aligned}\label{eqF8}
&\epsilon_{\hat{L}_{1r(2r)}} =\int_{0}^{t_g}\frac{dt}{T_{1}^{21}}\left[\frac{\sin^{2}(\Omega t)}{5}\right],\\
&\epsilon_{\hat{L}_{1\phi(2\phi)}} =\int_{0}^{t_g}\frac{2dt}{T_{\phi}^{21}}\left[\frac{\sin^{2}(\Omega t)}{5} -\frac{\sin^{4}(\Omega t)}{20} \right].
\end{aligned}
\end{equation}
Combining both contributions, the total incoherent gate error is
\begin{equation}
\begin{aligned}\label{eqF9}
\epsilon=\frac{1}{5}\frac{t_{g}}{T_{1}^{21}}+\frac{13}{40}\frac{t_{g}}{T_{\phi}^{21}}.
\end{aligned}
\end{equation}

\end{document}